\documentclass[structabstract]{aa}

\usepackage{graphicx}
\usepackage{epstopdf}
\usepackage[intlimits,sumlimits]{amsmath}
\usepackage{deluxetable}
\usepackage{times,epsfig} 
\usepackage{amssymb}
\usepackage{mathrsfs}  
\usepackage{natbib}

\bibpunct{(}{)}{;}{a}{}{,} 
\usepackage{caption}

\usepackage{amsmath}
\usepackage[english]{babel}
\usepackage{longtable}
\usepackage{url}
\usepackage{hyperref}

\newcommand{\beqa}{\begin{eqnarray}} 
\newcommand{\eeqa}{\end{eqnarray}}

\newcommand{\bsub}{\begin{subequations}}
\newcommand{\esub}{\end{subequations}}
\newcommand{\beal}{\begin{align}}
\newcommand{\ealn}{\end{align}}
\newcommand{\Nif}{$\rm ^{56}Ni$}

\def\V{SN~2006V}
\def\au{SN~2006au}

\newcommand{\Ni}{\ensuremath{^{56}\mathrm{Ni}}}
\newcommand{\Msun}{{\ensuremath{\mathrm{M}_{\odot}}}}
\newcommand{\Rsun}{{\ensuremath{\mathrm{R}_{\odot}}}}

\begin{document}

\title{The Type II supernovae 2006V and 2006au: two SN~1987A-like events\thanks{Based on observations
 collected at the European Organisation for Astronomical Research in the Southern Hemisphere, Chile (ESO Programme 076.A-0156).
This paper includes data gathered with the 6.5-m Magellan Telescopes located at Las Campanas Observatory, Chile.}}

\author{
F.~Taddia\inst{1} 
\and M.~D.~Stritzinger\inst{1,2}
\and J.~Sollerman\inst{1}
\and M.~M.~Phillips\inst{3}
\and J.~P.~Anderson\inst{4}
\and M.~Ergon\inst{1}
\and G.~Folatelli\inst{5}
\and C.~Fransson\inst{1}
\and W.~Freedman\inst{6}
\and M.~Hamuy\inst{4}
\and N.~Morrell\inst{3}
\and A.~Pastorello\inst{7,8}
\and S.~E.~Persson\inst{6}    
\and S.~Gonzalez\inst{3}
}

\institute{
Department of Astronomy, The Oskar Klein Center, Stockholm University, AlbaNova, 10691 Stockholm, Sweden 
\and 
{\it Affiliated member} Dark Cosmology Centre, Niels Bohr Institute, University of Copenhagen, Juliane Maries Vej 30, 2100 Copenhagen \O, Denmark 
\and 
Carnegie Observatories, Las Campanas Observatory, Casilla 601, La Serena, Chile 
\and 
Universidad de Chile, Departamento de Astronom\'{\i}a, Casilla 36-D, Santiago, Chile 
\and
Institute for the Physics and Mathematics of the Universe (IPMU), University of Tokyo, 5-1-5 Kashiwanoha, Kashiwa, Chiba 277-8583, Japan
\and 
Observatories of the Carnegie Institution for Science, 813 Santa Barbara St., Pasadena, CA 91101, USA
\and 
Astrophysics Research Centre, School of Mathematics and Physics, Queen University Belfast, Belfast BT7 1NN, United Kingdom
\and
INAF - Osservatorio Astronomico di Padova, Vicolo dell'Osservatorio 5, I-35122, Padova, Italy
}

\date{Received 14/09/2011 ; Accepted 25/10/2011}

\abstract
{Supernova 1987A revealed that a blue supergiant (BSG) star can end 
its life as a core-collapse supernova (SN). 
SN~1987A and other similar objects exhibit properties that distinguish 
them from ordinary Type II Plateau (IIP) SNe, 
whose progenitors are believed to be red supergiants (RSGs).
Similarities among 1987A-like events
include a long rise to maximum, early luminosity fainter than that of normal Type IIP SNe, and radioactivity acting as the primary source powering 
the light curves.} 
{We present and analyze two SNe monitored by the Carnegie Supernova Project 
that are reminiscent of SN~1987A.}
{Optical and near-infrared (NIR) light curves, 
and optical spectroscopy of SNe~2006V and 2006au are presented. 
These observations are compared to those of SN~1987A, 
and are used to estimate properties of their progenitors. }
{Both objects exhibit a slow rise to maximum and light curve 
evolution similar to that of SN~1987A. 
At the earliest epochs, SN~2006au also displays an initial dip 
which we interpret as the signature of the adiabatic cooling phase that ensues shock break-out.
SNe 2006V and 2006au are both found to be bluer, hotter and brighter than SN~1987A.
  Spectra of SNe~2006V and 2006au are similar to those of  SN~1987A and other 
normal Type II objects,
although both consistently exhibit expansion velocities higher than SN~1987A.
Semi-analytic models are fit to the UVOIR light curve of each object from which 
physical properties of the progenitors are estimated.  This yields ejecta mass estimates of
$M_{ej}\approx20$~$\Msun$, explosion energies of 
$E\approx2-3\times10^{51}$~erg, and progenitor radii of
$R\approx75-100$~$\Rsun$ for both SNe.} 
{The progenitors of SNe~2006V and 2006au were most likely BSGs with a larger explosion energy as compared to that of SN~1987A.}


\maketitle

\section{Introduction}

It is believed that hydrogen rich stars with masses larger than  $\sim8~\Msun$ end 
their lives as Type~II supernovae (SNe~II). 
Such supernovae typically reach peak brightness on time scales of 
days to a few weeks after explosion, and often this is followed by a 
``plateau" phase lasting up to several months.  
The plateau phase is produced by the recombination of 
hydrogen in the ejecta and provides the namesake to 
the sub-type designated as Type~IIP SNe (SNe~IIP). 
Both theory and observations
converge in asserting that the progenitors of SNe~IIP
are red supergiants (RSGs, see e.g. 
\citealp{grassberg71}). On the observational side,
this can be inferred from the length of the plateau phase 
 \citep{popov93}, and the progenitors themselves, having in a handful of cases
 been identified in pre-explosion images \citep[see][]{smartt09}.
  
The best ever studied supernova  -- SN~1987A -- showed  
that blue supergiants (BSGs) also produce SNe~IIP, although with somewhat peculiar light curves.
In this case the early bolometric luminosity was considerably smaller
than for a typical SN~IIP, suggesting that the majority of the explosion 
energy was used to adiabatically expand the compact BSG star.
The light curves of SN~1987A displayed an unusually long rise to maximum 
and were mainly powered by the radioactive decay of $\Ni$ and its daughter product $^{56}$Co (see
\citealp{arnett89} for a review on SN~1987A,
including a discussion on its progenitor).

Given the impact of SN~1987A on our general understanding of 
SN physics, identifying additional objects which have BSG progenitors 
is of importance.
Following the discovery of SN~1987A and its association with a BSG, 
11 years passed before another similar SN was found.  
Like its predecessor, SN~1998A  displayed a $\sim$~100 day rise 
to maximum. It also exhibited higher expansion velocities and was more 
luminous \citep{woodings98,pastorello05} than SN~1987A.  
\citet{pastorello05} argued that the progenitor of SN~1998A was a BSG. 
 \citet{kleiser11} recently published a study  
of two other objects, SNe~2000cb and 2005ci,
that may also have a BSG origin, and most recently,  
\citet{pastorello11} presented observations of SN~2009E which they argue  
to have a BSG progenitor as well. 

In this paper we present observations obtained within the
{\em Carnegie Supernova Project} (CSP; \citealp{hamuy06}) of 
the 1987A-like SNe~2006V and 2006au.
 We investigate the properties of these two SNe and 
compare their photometric and spectroscopic characteristics to
SN~1987A and other peculiar SNe~II. 
We find that these two objects have similar light curve 
shapes and spectral properties as SN~1987A.  
In what follows our data set is used
to  place limits on the  physical parameters of their 
compact progenitor stars. 
From this analysis we suggest the progenitors of these SNe were BSGs. 

\begin{figure}
\centering
\includegraphics[width=6in,height=6.0in]{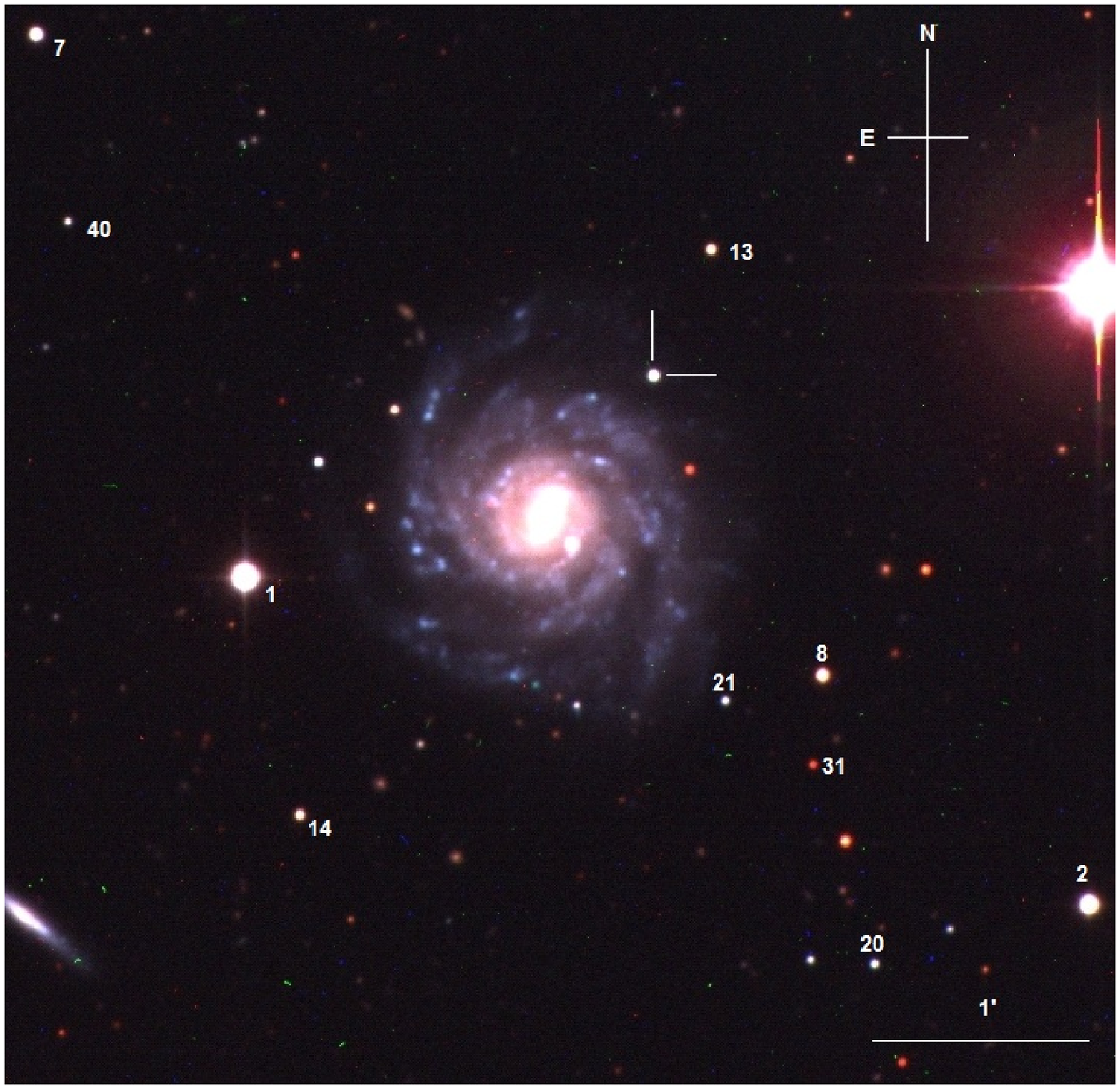}
\caption{\label{gal06V}A composite optical image of UGC~6510 with the position of SN~2006V indicated, along with the orientation, the scale and some local sequence stars. The optical images were taken with the Swope telescope.}
\end{figure}

\V\ was discovered on 4.7 February 2006 UT \citep{chen06}
 in the SAB(rs) galaxy UGC~6510 
during the course of the Taiwan
Supernova Survey.  
With J2000.0 coordinates $\alpha = 11^{\rm h}31^{\rm m}30.01^{\rm s}$, $\delta =
-01^\circ17\arcmin52\farcs2$, the SN was positioned 
32\farcs1 West and 40\farcs4 North from the host galaxy center (see Fig.~\ref{gal06V}).  
On 7.5 February UT, \citet{blondin06} 
spectroscopically classified SN~2006V
as a Type~II SN, and noted the absorption in H$\alpha$ to be
blue-shifted by roughly 4000~km~s$^{-1}$. 
According to the Nasa Extragalactic Database (NED),  
UGC~6510 has a heliocentric redshift $z=0.0158$.  
From the narrow H$\alpha$ emission 
line measured in 
the optical spectra presented below, 
we measure a value
of $z=0.0157\pm 0.0013$ (the error corresponds to an uncertainty of $\pm 400$~km~s$^{-1}$).
Adopting a Hubble constant of  
H$_0$ =  $73.8\pm2.4$~km~s$^{-1}$~Mpc$^{-1}$ 
\citep{riess11}
our measured redshift from the spectra of 
SN~2006V corresponds to a distance of $72.7\pm5.0$~Mpc. Here
we have taken into account 
peculiar motion corrections (Virgo+GA+Shapley, see Mould et al. 2000) 
and  cosmological parameters of  $\Omega_{m}=0.30$ and $\Omega_{\Lambda}=0.70$.

\begin{figure}
\centering
\includegraphics[width=6in,height=6.0in]{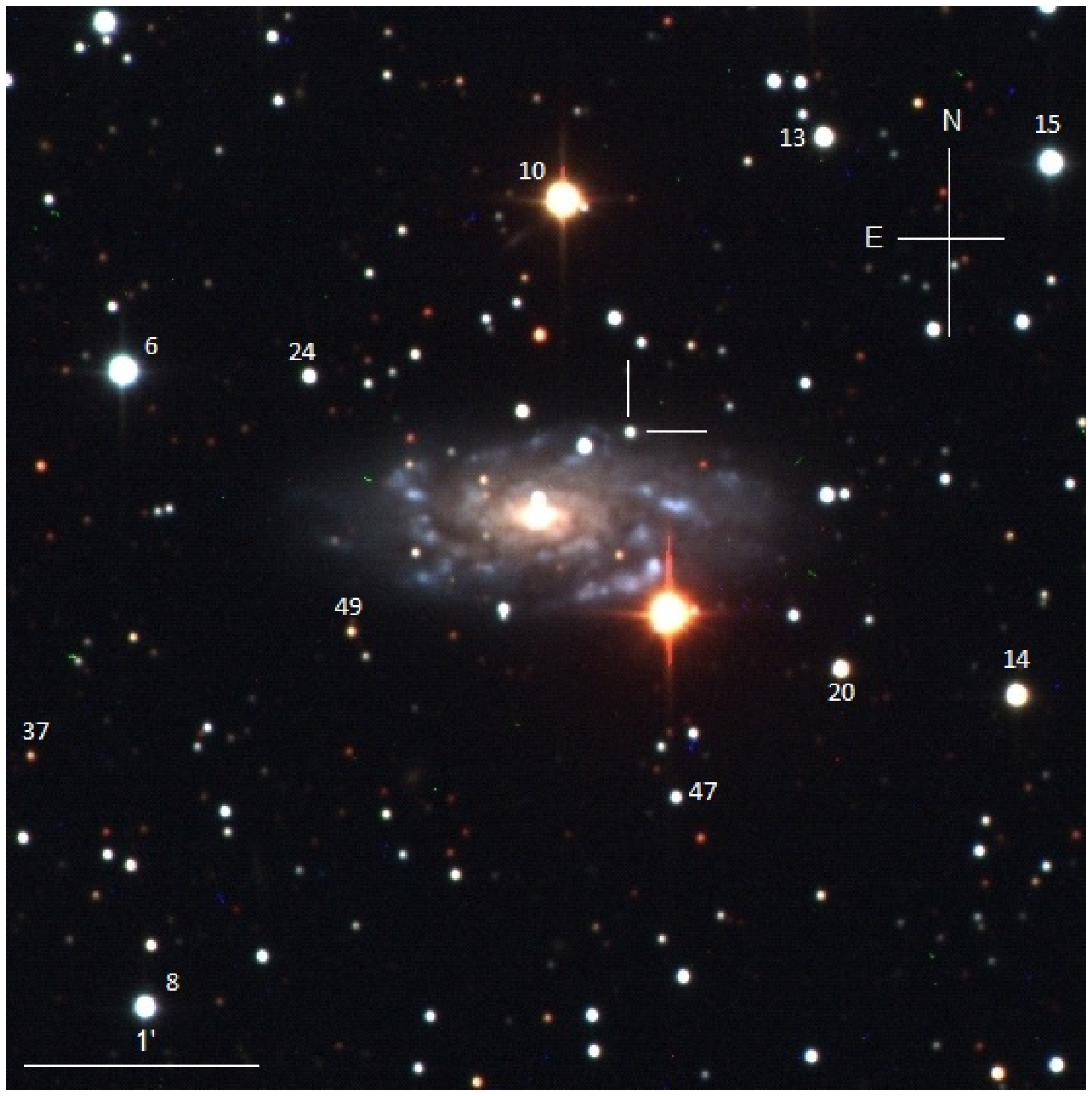}
\caption{\label{gal06au} A composite optical image of 
UGC~11057 with the position of SN~2006au indicated, along with the orientation, the scale and some local sequence stars. The optical images were taken with the Swope telescope.}
\end{figure}

\au\  was discovered on 7.2 March 2006 UT  
in the Scd galaxy UGC~11057
by the Tenagra Observatory Supernova Search \citep{trondal06}.  
With J2000.0 coordinates 
 $\alpha = 17^{\rm h}57^{\rm m}13.56^{\rm s}$, $\delta =
12^\circ11\arcmin03\farcs2$, this object was located 
17\farcs0 West and 18\farcs2 North from the center of the host galaxy (see Fig.~\ref{gal06au}).
On 13.6 March UT, \citet{blanc06} classified SN~2006au
as a Type~II SN, and noted a well developed H$\alpha$ P-Cygni profile. 
NED lists a heliocentric redshift to UGC~11057 of $z = 0.0099$, which 
is in agreement with the value measured from our 
spectra of $z = 0.0098\pm0.0013$.
Our measured redshift to UGC~11057 
corresponds to a distance of $46.2\pm3.2$~Mpc, 
where we again adopted the above-mentioned
cosmology and peculiar motions.

The organization of this paper is  as follows:
Section~\ref{sec:data_red} contains brief details regarding the 
observations and subsequent data reduction techniques;
Section~\ref{sec:photometry} presents the 
broad-band optical and near-infrared photometry;
Section~\ref{sec:spectra} contains the 
spectroscopic observations; Section~\ref{sec:discussion} gives
 our discussion including a comparison
with other 1987A-like objects and SNe from BSGs; and this is
 followed by our conclusions in Section~\ref{sec:conclusions}.

\section{Data Acquisition and Reduction}
\label{sec:data_red}

Broad-band imaging of SNe~2006V and 2006au  was  obtained
with facilities  at the Las Campanas Observatory (LCO). 
Optical ($ugriBV$) and near-infrared ($YJH$) 
imaging was performed with the Henrietta Swope 1-m 
and the Ir\'en\'ee du Pont 2.5-m telescopes equipped with CSP filters. 
The optical images were obtained with the Swope Direct Camera equipped with the CCD named Site 3,
 and with the du Pont Direct Camera equipped with the CCD named Tek 5.
In the following we adopt Site 3 and Tek 5 as names to distinguish these cameras.
Near-infrared imaging was obtained with RetroCam on Swope and with the Wide Field IR Camera (WIRC) on the du Pont. 
Details regarding  these instruments  and the bandpasses used are given in \citet{hamuy06},
 \citet{contreras10}, and \citet{stritzinger11}.

Detailed descriptions of the observing techniques and 
data reduction methodology can be found in \citet{contreras10}; 
in what follows we briefly summarize the data reduction process.
All optical images were reduced in a standard manner including:
(i) bias subtraction, (ii) flat-field division, and (iii) the application of
a shutter time and linearity correction. 
The near-infrared images were also reduced following several steps,
consisting of (i) dark subtraction, (ii) flat-field division, (iii) sky
subtraction, and (iv) geometric alignment and combination of the
dithered frames. 

Months after each SN faded, deep template images of their host galaxy 
were obtained under excellent seeing conditions. 
Optical and near-infrared template imaging was performed 
with the du Pont telescope using Tek 5 and WIRC.
Following the method highlighted in \citet{contreras10},
the template images allow us to subtract away the host background 
light at the position of the SN in each science image.

 Observed magnitudes of each SN were computed differentially from the science frames with respect 
 to a local sequence of stars. The photometric sequences were calibrated using
 \citet{landolt92} ($BV$), \citet{smith02} ($ugri$)  
 and \citet{persson98} ($YJH$) fields observed over a minimum of 
 three photometric nights. 
The local sequences of SNe~2006V and 2006au in the 
 standard system are
 provided in Table~\ref{tabstars}.

Nine epochs of optical spectroscopy were obtained for both SNe~2006V and 2006au
with telescopes at LCO, and in one instance each, with the New Technology Telescope (NTT). 
A journal of the spectroscopic observations is provided in  
Table~\ref{tabspectra}.  Depending on the exact instrument used, the
wavelength interval generally ranges from $\sim$~3800 \AA\ to 9500 \AA.
Standard reductions of each spectrum were performed as described in 
\citet{hamuy06}. 
Briefly, this consisted of overscan correction, bias subtraction and flat-fielding. 
The 2-D spectra were then optimally extracted and wavelength calibrated with respect 
to arc lamps. The wavelength corrected spectra were corrected for telluric
absorption through the division of a telluric standard spectrum, and then
flux-calibrated. Multiple exposures of a particular object were then combined to produce
a final high signal-to-noise science product. This allowed for the removal of cosmic rays.

\section{Photometry}\label{sec:photometry}

\begin{figure}
\centering
\includegraphics[width=6in]{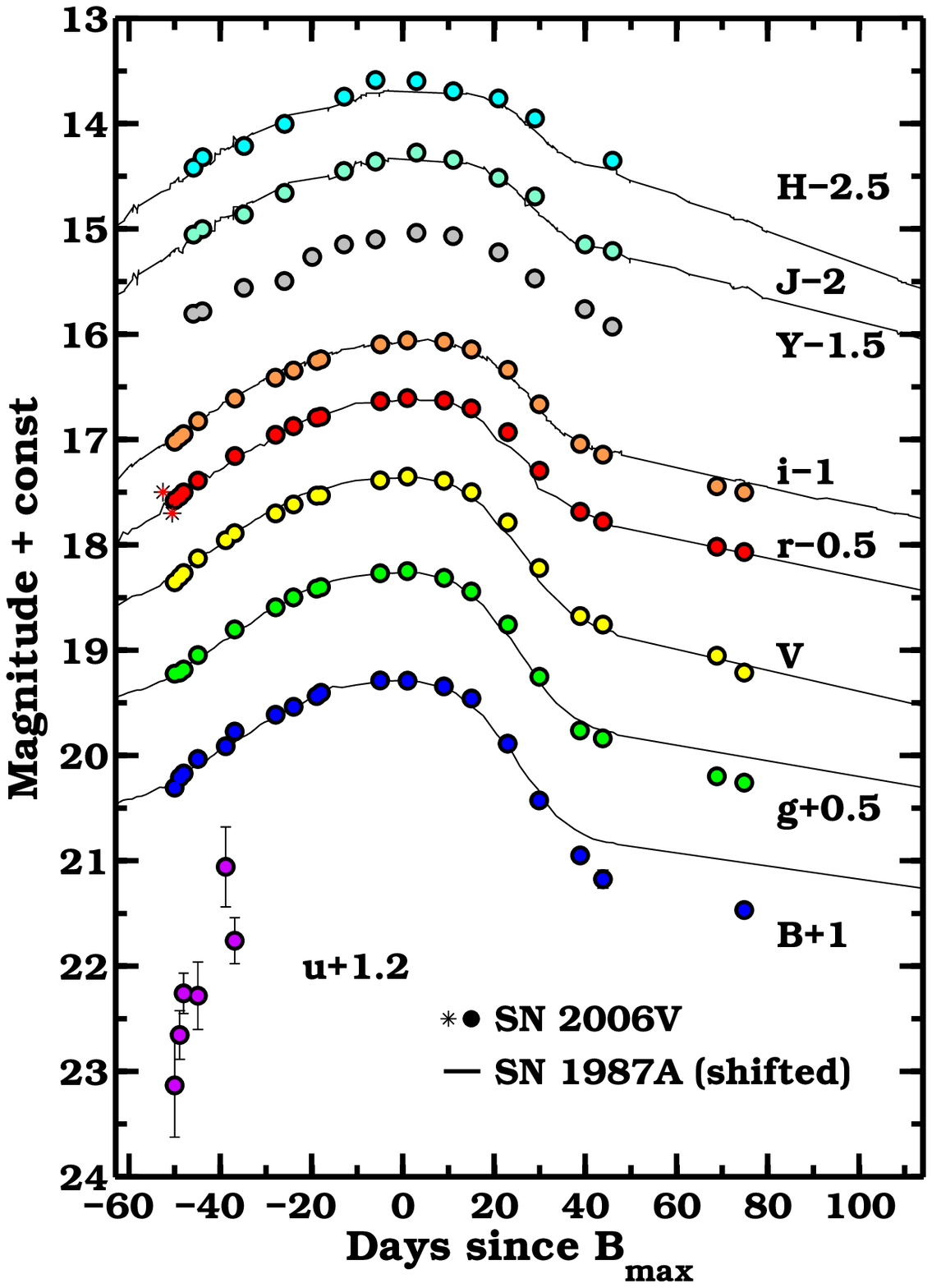}
\caption{\label{lc06V} 
 Optical and near-infrared light curves of \V. 
Over-plotted as solid lines are light curves of SN~1987A ($J$ and $H$ from \citealp{bouchet89}; $I$ from \citealp{hamuy88}; $BgVr$ from spectrophotometry on the spectra sample published in \citealp{phillips88}) scaled to match the peak brightness of SN~2006V. We also show (stars) the unfiltered/$r$ magnitudes from discovery and confirmation images \citep{chen06}. The similarity of the light curve shape
of these two objects is striking.}
\end{figure}

\begin{figure}
\centering
\includegraphics[width=6in]{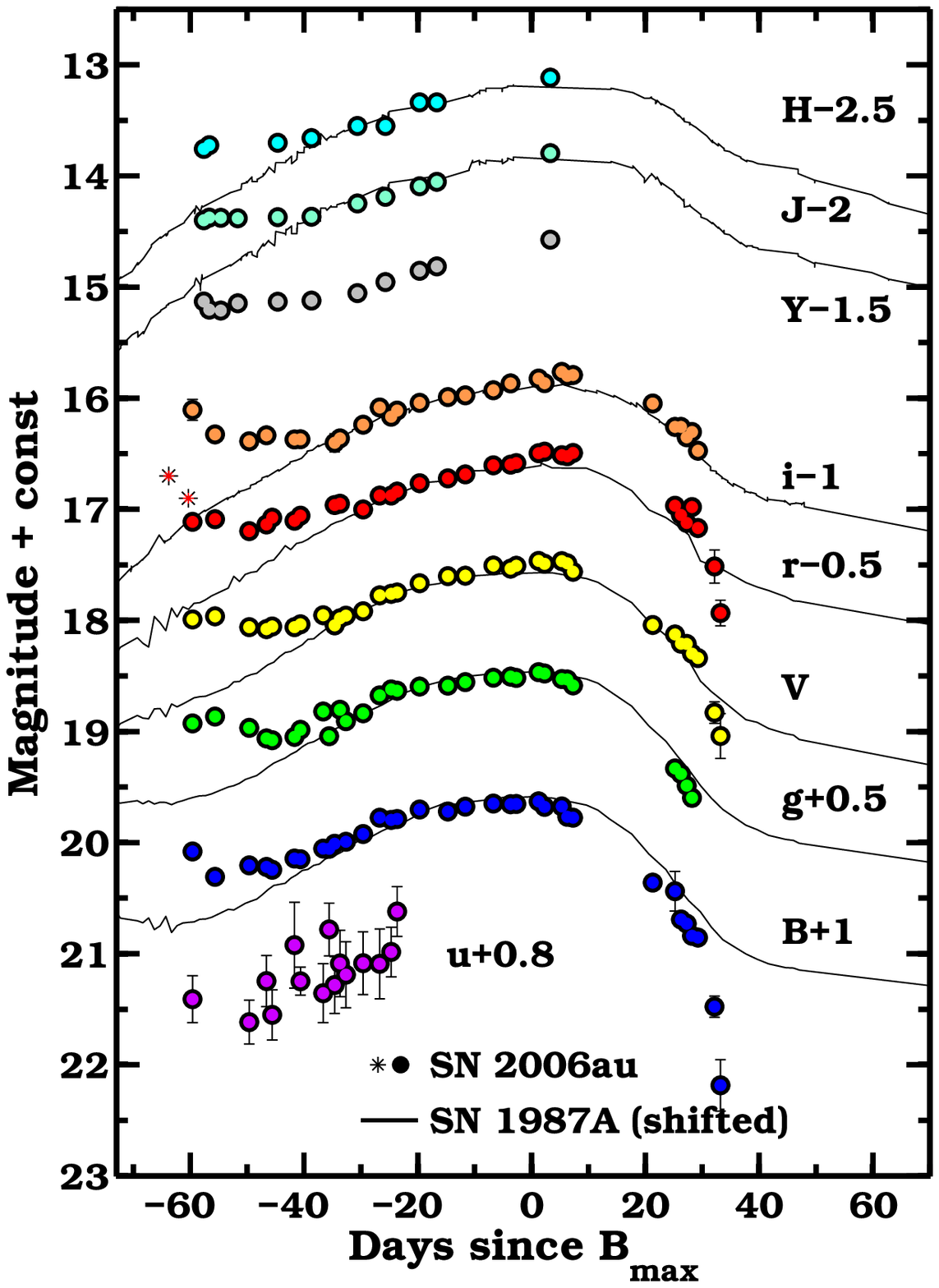}
\caption{\label{lc06au} Optical and near-infrared light curves of \au. Over-plotted as solid lines 
are light curves of SN~1987A ($J$ and $H$ from \citealp{bouchet89}; $I$ from \citealp{hamuy88}; $BgVr$ from spectrophotometry on the spectra sample published in \citealp{phillips88}) scaled to match the peak brightness of SN~2006au. We also show (stars) the unfiltered/$r$ magnitudes from discovery and confirmation images \citep{trondal06}.
The two objects have nearly identical evolution between $-30$ to $+$30 days past maximum. 
At earlier epochs \au\ shows evidence of the tail end of the cooling phase 
that followed shock break-out. At later epochs the $B$, $V$ and $r$-band light curves of 
SN~2006au appear to drop at a faster rate.}
\end{figure}

Optical and near-infrared light curves of SNe~2006V and 2006au 
are plotted in Figs.~\ref{lc06V} and \ref{lc06au}, respectively. 
The corresponding
optical photometry
-- in the CSP natural system --
is listed in Tables~\ref{tabphot06V} and \ref{tabphot06au}, whereas the final near-IR photometry 
in the Persson et al. (1998) system is listed in Tables~\ref{tabphotIR06V} and \ref{tabphotIR06au}.
The differences in measured magnitudes between the natural and the standard system are insignificant for the purpose
 of this study \citep{hamuy06}, whereas they would be important for precision cosmology. The transformation equations
 between the two systems are reported in \citet{hamuy06}, with updated color terms provided by \citet{stritzinger11}.
 The light curves of  SN~2006V follow the flux evolution from  $-50$ to $+75$ days 
past $B$-band maximum ($B_{\rm max}$, i.e. $JD=2453823.7$), while those of 
SN~2006au range from $-59$ to $+33$ days past $B_{\rm max}$ ($JD=2453865.5$).  

The long rise to maximum, the
broad peak and the subsequent decline to the radioactive tail are very
similar to those shown by the light curve of SN~1987A.
This is evident in Figs.~\ref{lc06V} and \ref{lc06au}, where we have also 
included for comparison the light curves of SN~1987A as 
thin solid lines. 
In particular, we show $U$- and $I$-band from \citet{hamuy88} and $J$- and $H$-band from \citet{bouchet89}. 
$B$, $g$, $V$ and $r$ light curves of SN~1987A are synthetic magnitudes, which have been computed using CSP bandpasses and 
  spectrophotometry published by \citet{phillips88}. 

Although both SNe 2006V and 2006au were spectroscopically 
classified as SNe~II (\citealp{blanc06}; \citealp{blondin06})
their sustained rise to, and 
evolution through maximum, suggests they are more appropriately 
termed peculiar 1987A-like SNe  \cite[see also][]{pastorello11}.

Taking a closer look at the early phase evolution of \au,
reveals that the light curves first decrease in brightness 
on a time-scale of $\sim~2-3$ weeks. 
This is most evident 
in the $B$, $i$ and $r$ bands (the unfiltered/$r$ magnitudes from discovery and confirmation images, \citealp{trondal06},
 confirm this trend), followed to a lesser extent in the $gV$ bands 
while the evolution in $YJH$ is nearly flat.

A similar evolution was observed in the $U$- and $B$-band 
light curves of SN~1987A \citep{hamuy88}, as well 
as in other core-collapse SNe caught just after explosion;
e.g. SNe 1993J \citep{richmond94},
1999ex \citep{stritzinger02}, 2008D \citep{malesani09,modjaz09} and 2011dh \citep{soder11}.
For these objects, that were all caught very early, this
light curve evolution was interpreted as a sign of the 
photospheric cooling phase that ensues shock-wave breakout.
The decreasing phase is not as evident in SN 2006au, 
but we propose a similar scenario for this supernova 
although its somewhat later discovery (see subsection \ref{date}) does not
allow to see a clearly decreasing temperature at early epochs
(see middle panel in Fig.~\ref{bolo}).

The last photometric observations of SNe 2006V and 2006au were obtained
at $+75$ and $+33$ days past $B_{\rm max}$, respectively. For this reason,
 we can only observe the linear decay phase for \V, whose optical light curves
settle onto a similar radioactive decay slope as did SN~1987A. 
For \au\ no later phase data are 
available to map out the linear decline phase.
However,
we do note that at epochs later than
+30 days past $B_{\rm max}$, the evolution appears to be faster 
than in the case of SN~1987A.

In order to estimate the absolute magnitudes and  
luminosities of SNe~2006V and 2006au, accurate estimates
of Galactic reddening and reddening associated with dust in the host galaxies are needed.
In the case of SN~2006V, NED lists a Galactic color excess value of 
$E(B-V)_{\rm MW} = 0.029$ mag \citep{schlegel98}. 
Close examination of our spectroscopic sequence
of this object 
(Sect.~\ref{sec:spectra}) shows no evidence for \ion{Na}{i}~D absorption.
The presence of \ion{Na}{i}~D is often considered to be a proxy for dust attenuation, 
therefore the lack 
of \ion{Na}{i}~D combined with 
the position of SN~2006V in the outskirts of its face-on galaxy suggests 
minimal 
host extinction.
In the following we therefore assume zero
host extinction for SN~2006V. Adopting the standard total-to-selective
extinction value of $R_V = 3.1$ \citep{cardelli89}
the total color excess in the direction of SN~2006V corresponds to 
a modest  $A_{V} = 0.09$~mag. 

The Galactic color excess in the direction of SN~2006au is 
$E(B-V)_{\rm MW} = 0.172$~mag \citep{schlegel98}.   
Conspicuous \ion{Na}{i}~D absorption lines 
at the redshift of UGC~11057 are detected in the first spectrum of SN~2006au.
This is shown in the inset of Fig.~\ref{spectra06au}.
Two Gaussian profiles were fit to the (unresolved) \ion{Na}{i}~D lines 
and a total equivalent width of 0.88$\pm$0.11~\AA\
is measured. 
Using the \citet{turatto03} correlation
between the \ion{Na}{i}~D equivalent width and host galaxy reddening  
suggests a host galaxy color excesses $E(B-V)_{host}=0.141$~mag.  
This value might be affected by large uncertainty and lead to 
an overestimated host extinction, since we know that the \citet{turatto03}
relation presents large scatter \cite[see][]{poznanski11}. 
However, we combined this value with the Galactic color excess, obtaining
$E(B-V)_{tot}=0.312$~mag, which corresponds to  
 $A_{V}=0.97$~mag.

\begin{figure}
\centering
\includegraphics[width=6in]{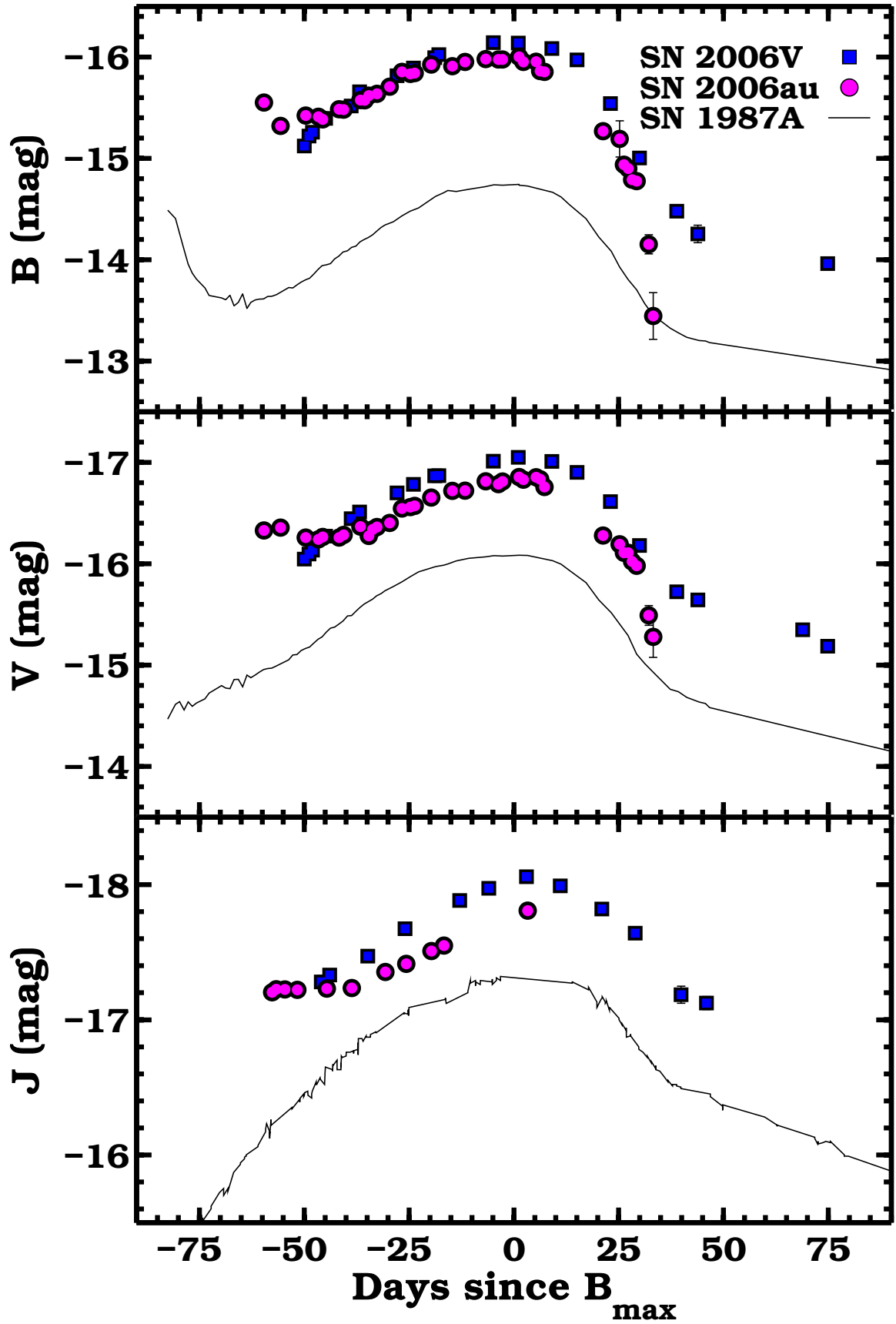}
\caption{\label{absb}Absolute $B$ (top panel), $V$ (middle panel) and $J$ (bottom panel) light curves of 
SNe~1987A \citep{hamuy88}, 2006V and 2006au.
For SN~1987A  we adopt $E(B-V)_{tot}=0.175$~mag \citep{woosley87} and a distance of $d=50\pm5.2$~kpc \citep{storm04}.}
\end{figure}

Armed with estimates of $E(B-V)_{tot}$, the absolute magnitudes of 
SNe~2006V and 2006au are computed for each observed bandpass.
Their peak values are given in Table~\ref{absphotPEAK}. 
Compared to SN~1987A, SNe~2006V and 2006au are
brighter in all bands.
To demonstrate the differences, in Fig.~\ref{absb} 
the absolute $B$-, $V$- and $J$-band light curves of SNe~2006au and 2006V 
are compared to those of SN~1987A.
SNe~2006V and 2006au show 
an absolute $B$ magnitude difference to SN~1987A of $\sim$~1.5 mag and $\sim$~1.3 mag respectively. 
The differences in the $V$ band are smaller ($\sim$~1.0 and $\sim$~0.7 mag) and
in the $J$ band are further reduced ($\sim$~0.8 and $\sim$~0.5 mag).

\begin{figure}
\centering
\includegraphics[width=6in]{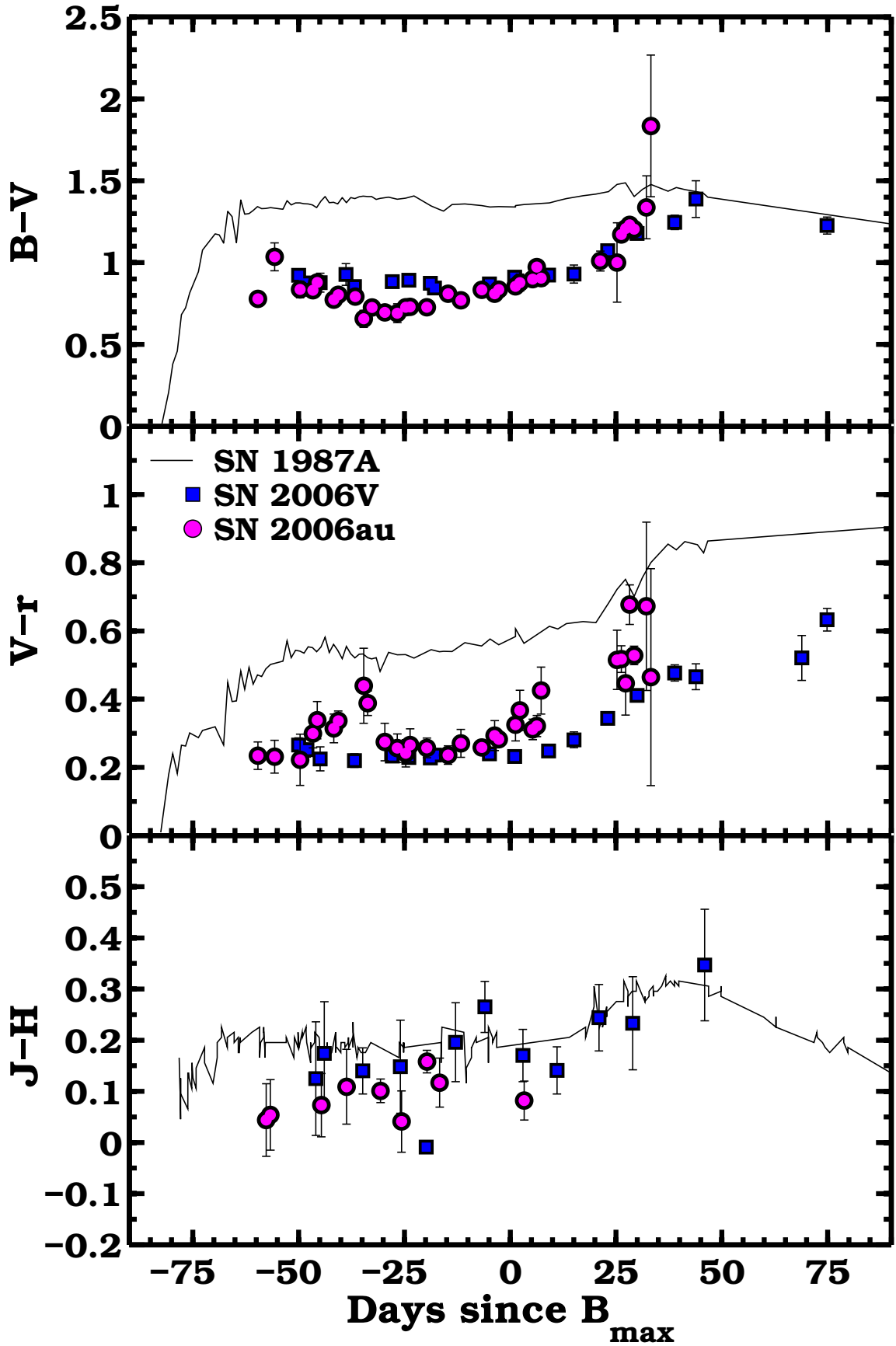}
\caption{\label{color}$B-V$ (top panel), $V-r$ (middle panel) and $J-H$ (bottom panel) color curves 
for SNe~2006V, 2006au, and 1987A. Both SNe~2006V and 2006au are considerably bluer than SN~1987A.}
\end{figure}

To gain insight into the photospheric temperatures, we plot in 
Fig.~\ref{color}  
the $B-V$, $V-r$ and $J-H$ color curves of SNe~2006V and 2006au. 
The plot also includes the colors of SN~1987A. The photometry of each SN has been corrected for extinction 
adopting the values previously mentioned. Overall the optical colors
 of SNe~2006V and 2006au are bluer than for SN~1987A, particularly during the earliest epochs
 where the $B-V$ (top panel) and $V-r$ (middle panel) color differences amount to $\sim$~0.7~mag and $\sim$~0.3~mag, respectively. 
 Later, around $B_{\rm max}$, the $B-V$ colors of our two objects evolve towards the 
red, and are at day $+$35 comparable to the $B-V$ color 
of SN~1987A. The $V-r$ color of \au\ follows a similar 
evolution (at the earliest epoch it is bluer than SN~1987A by $\sim$~0.3~mag). 
In contrast, the $V-r$ color of 
SN~2006V is bluer than SN~1987A at all epochs. 
The near-infrared color $J-H$ (bottom panel) of 
SN~2006V is found to be similar to SN~1987A at all epochs, 
whereas for SN~2006au it is slightly bluer at the 
earliest phase (by $\sim$~0.1~mag).

The color comparisons suggest that the photospheres of 
SNe~2006V and 2006au are at higher temperatures than
for SN~1987A. This is confirmed by  black-body fits to the spectral energy distributions 
(SEDs) obtained from each photometric epoch; 
see middle panel in Fig.~\ref{bolo}.
As shown in this plot, the black-body temperatures of SNe~2006V and 
2006au are higher than for SN~1987A. 
We excluded $u$ and $B$ from the fits, 
given their significant deviation from the Planck function.

\section{Spectroscopy}\label{sec:spectra}

\begin{figure}
\centering
\includegraphics[width=6.0in]{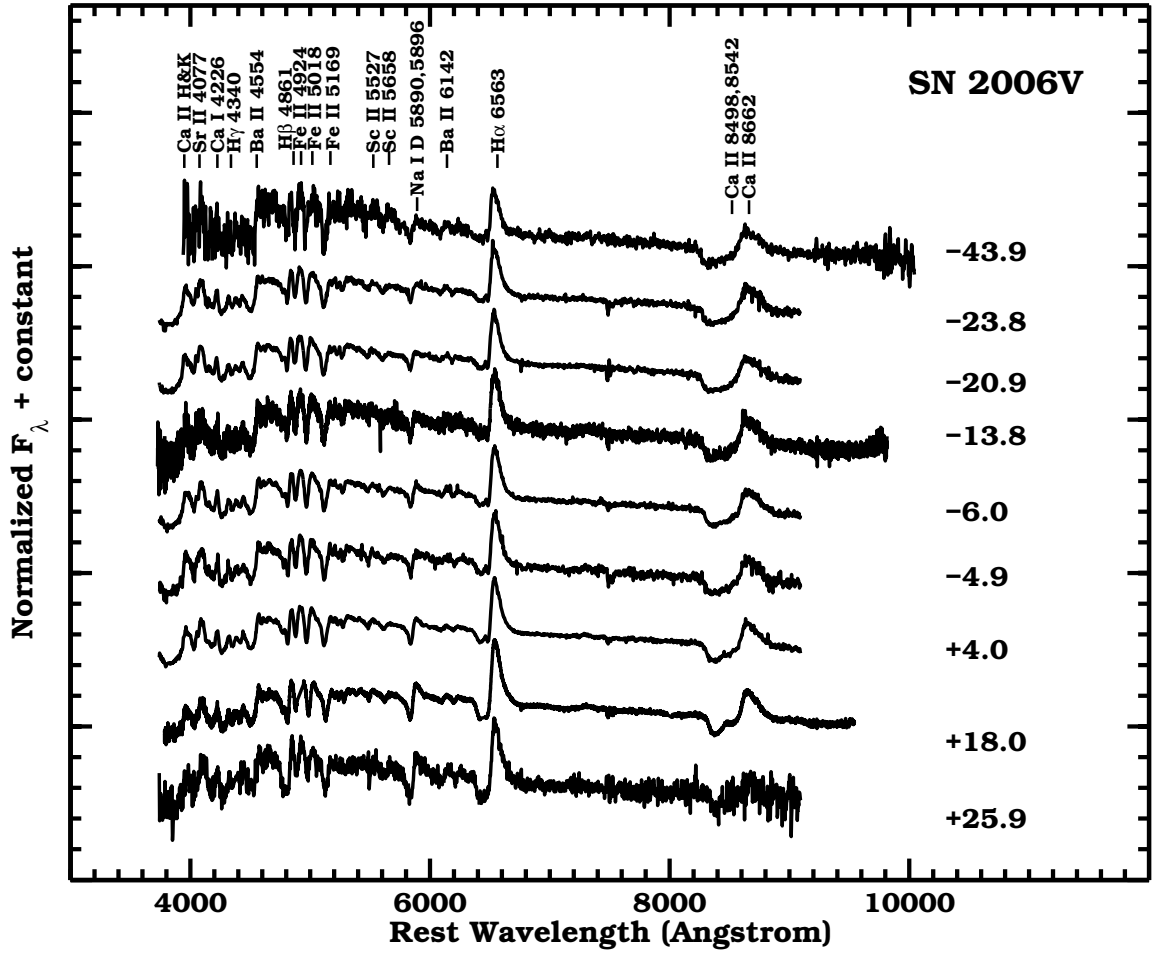}
\caption{\label{spectra06V}Spectral evolution of \V. Days since $B_{\rm max}$ are 
reported in the figure. The spectra exhibit
strong H$\alpha$ P-Cygni profiles, 
\ion{Fe}{ii} features and \ion{Na}{i}, and are similar to those of 
normal SNe~II.}
\end{figure}

\begin{figure}
\centering
\includegraphics[width=6.0in]{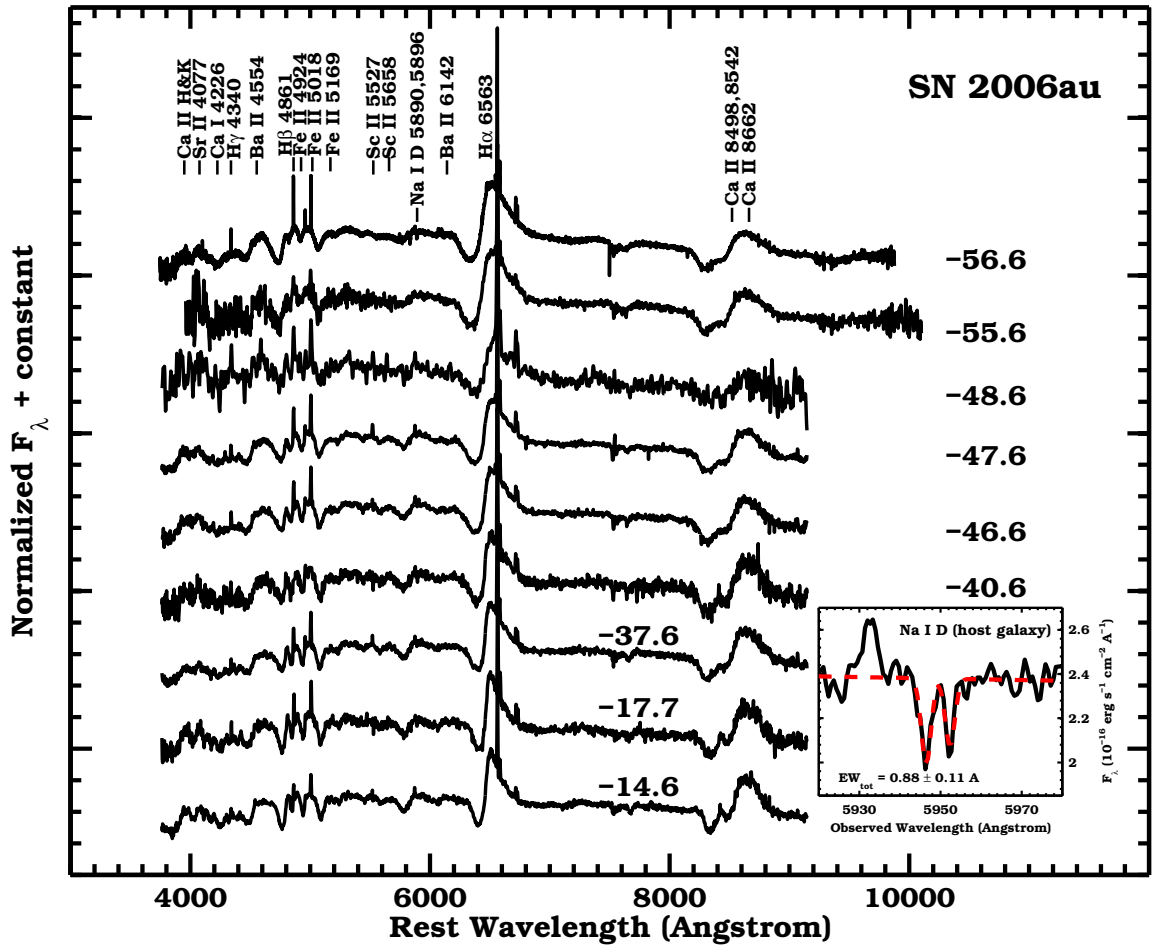}
\caption{\label{spectra06au}Spectral evolution of \au. Days since $B_{\rm max}$ are reported in the figure. 
Overall these spectra resemble those of \V. The narrow emission lines are from the host galaxy. Shown in the inset is the spectrum obtained on 
14th March 2006 plotted in the rest frame of the host galaxy, in the wavelength range where narrow \ion{Na}{I}~D absorption feature has been detected. Over-plotted (red dashed line) is the best fit 
of two Gaussians with a FWHM of 2.8~\AA, from 
which we obtain a total equivalent width of $0.88\pm0.11$~\AA.}
\end{figure}

Spectroscopic sequences of SNe~2006V and 2006au are plotted in 
Figs.~\ref{spectra06V} and \ref{spectra06au}, respectively. The spectra of SN~2006V 
cover the flux evolution from day $-44$ to $+25$ relative to 
$B_{\rm max}$, and for SN~2006au from day $-56$ to $-14$. 
The spectra of both objects exhibit common features typically 
observed in SNe~II.

The earliest spectra of SN~2006V show all the typical features of a normal SN~IIP
at one month after the explosion. Together with
spectral lines that are expected to be visible already at early phases
(H$\alpha$, H$\beta$, Ca~H$\&$K and the NIR triplet), the \ion{Na}{i} doublet and
\ion{Fe}{ii} lines become prominent,  particularly those of the 
\ion{Fe}{ii} multiplet 42.

Later on, weak lines of \ion{Sc}{ii}  (5527~\AA, 5641-5669~\AA, 6246~\AA) and the
unblended \ion{Ba}{ii}  $\lambda$6162 are clearly visible. The subsequent spectra of 
SN~2006V show a modest evolution with the continuum becoming marginally
redder with time. Strong line blanketing can be observed below $\sim$4400
\AA. This is probably due to the increasing strength of lines of 
\ion{Cr}{ii}, \ion{Ti}{ii}, \ion{Fe}{i}.
In post-maximum spectra of SN~2006V a hint of the nebular \ion{Ca}{ii}
7291-7323~\AA\ is seen.

The spectra of SN~2006au are contaminated by the host galaxy background.
A number of narrow lines due to underlying \ion{H}{ii} regions are 
identified including:
H$\alpha$, H$\beta$, [\ion{O}{iii}]. 
Nevertheless, the spectra of SN~2006au
show an evident evolution, with \ion{Na}{i}, \ion{Sc}{ii} and \ion{Ba}{ii} lines becoming
more prominent with time. This is in good agreement with the fact that the
first spectrum for SN~2006au was taken at an earlier phase than for SN~2006V.
Unfortunately post-maximum spectra showing nebular features are not available for SN~2006au.
 
\begin{figure}
\centering
\includegraphics[width=6.0in]{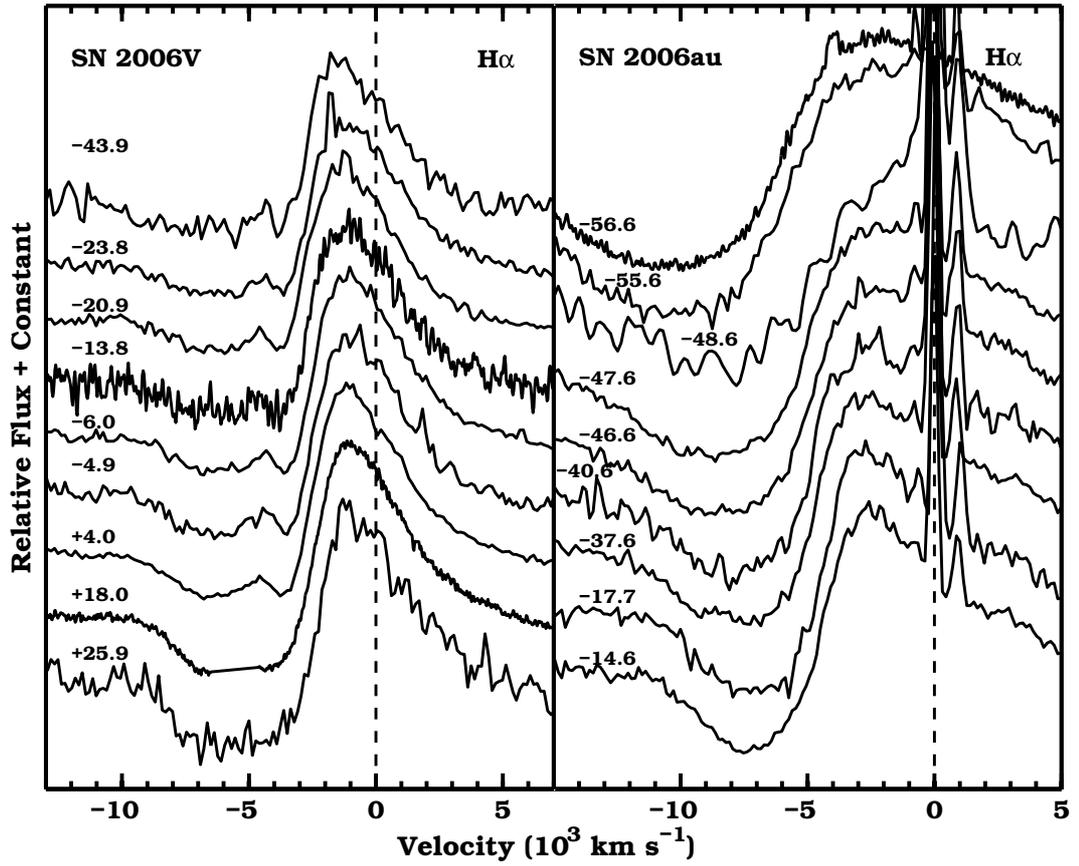}
\caption{\label{Halpha}H$\alpha$ P-Cygni profile evolution for \V\ (left panel) and \au\ (right panel). 
Since the spectra of \V\ are not particularly early (days since $B_{\rm max}$ are reported on the left), the velocity corresponding to the minimum of absorption is almost constant. For \au, the shift towards lower velocities in the last six spectra is evident.}
\end{figure}

\begin{figure}
\centering
\includegraphics[width=6.0in]{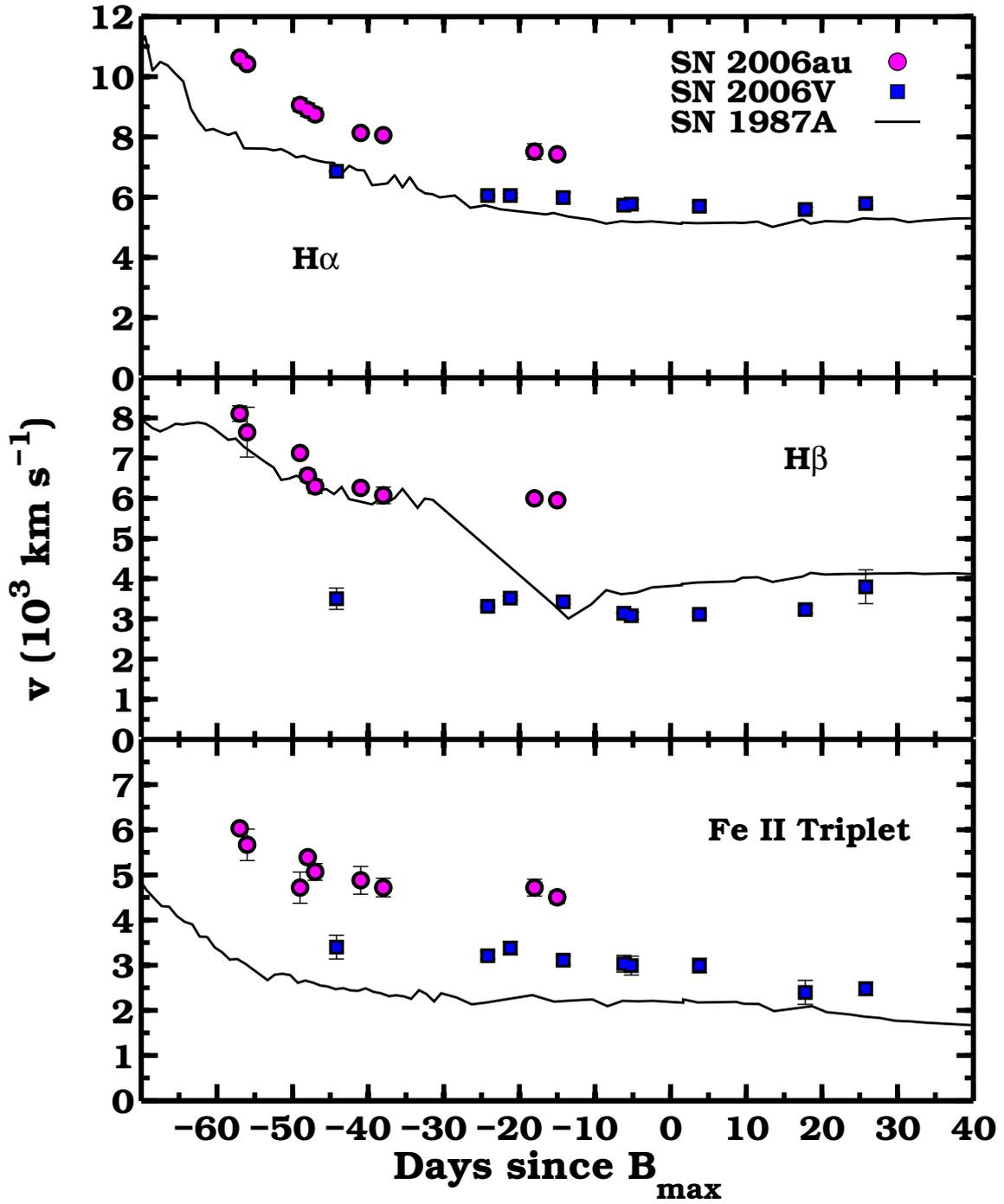}
\caption{\label{velocity}Measured expansion velocities for SNe 2006V, 2006au and 1987A, from the minimum of absorption of H$\alpha$ (top panel), H$\beta$ (middle panel) and \ion{Fe}{ii} multiplet 42 (bottom panel).}
\end{figure}

Plotted in Fig.~\ref{Halpha} 
are the H$\alpha$ profiles as a function of time. 
In the case of  \au\
the minimum of the absorption clearly evolves towards lower velocities, 
while
in SN~2006V, where the spectra were obtained at later epochs,
the H$\alpha$ velocity is nearly constant at $\sim$~6000~km~s$^{-1}$.
This is quantitatively shown in Fig.~\ref{velocity} where 
the expansion velocities
of  H$\alpha$  (top panel), H$\beta$ (middle panel) and the \ion{Fe}{ii} multiplet 42 
(bottom panel) are measured from the minimum of their respective 
 absorption features. Also included in this figure are the expansion velocities 
 of SN~1987A measured from spectra published by \citet{phillips88}. 
 Following  \citet{meaburn95}, the spectra of SN~1987A were 
corrected for a redshift of $\Delta v=286$~km~s$^{-1}$.
The expansion velocities of SN~2006au are clearly 
higher than for both of the other two objects, while SN~2006V 
shows marginally higher velocities than SN~1987A at nearly all epochs when we look
at the \ion{Fe}{ii} lines, which are indicative of the photospheric velocity. 

\begin{figure}
\centering
\includegraphics[width=6.0in]{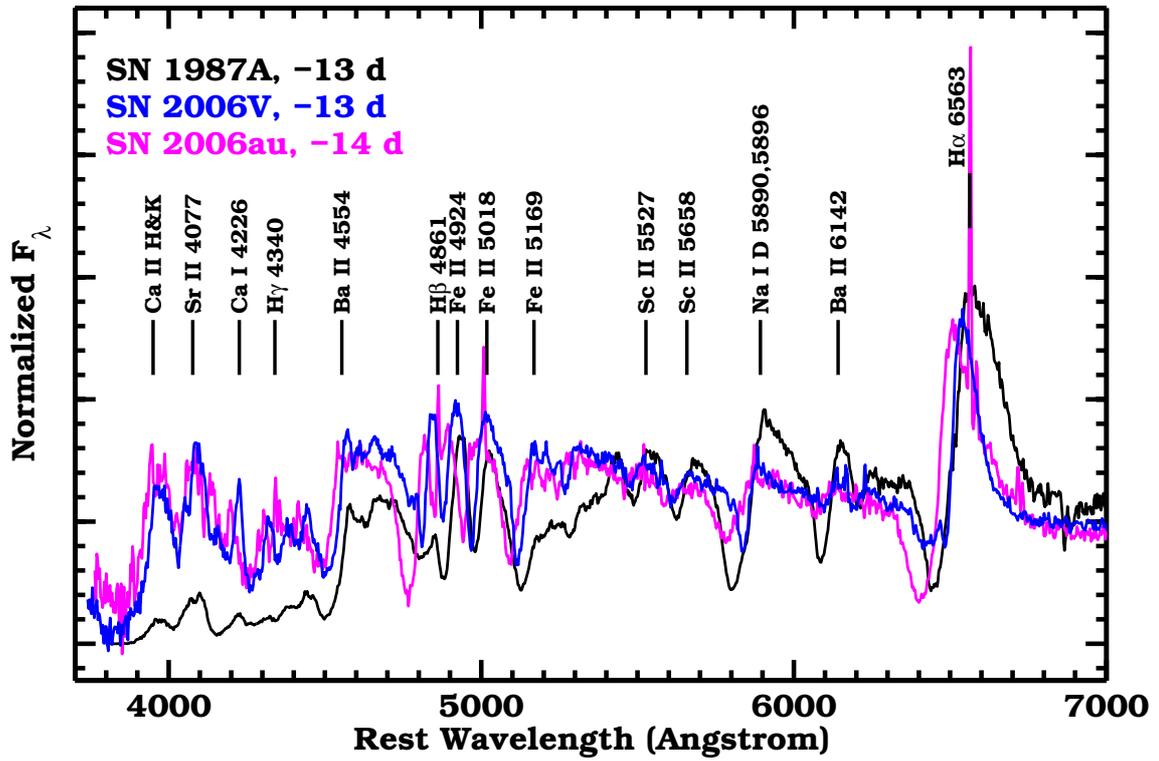}
\caption{\label{spectracomp} Optical spectral comparison of SNe 2006au, 2006V and~1987A, about two weeks before $B_{\rm max}$. Both 
SNe~2006V and 2006au appear bluer, but the spectral features are comparable, although the intensities are different.}
\end{figure}

Finally, in Fig.~\ref{spectracomp} we compare spectra taken 2 weeks before 
maximum of SNe~1987A, 2006V and 2006au.  
Each spectrum has been corrected for extinction and redshift. 
In line with the color evolution, the spectra of SNe~2006V 
and 2006au appear to be bluer than those of SN~1987A. The blue region of 
the spectra is heavily suppressed in SN~1987A, whereas
for SN~2006V and SN~2006au the flux is not much lower than in the red region.
Indeed, for our objects, the shape of the spectral continuum is well
 approximated by a black body function of relatively high ($\sim$6000~K) 
 and almost constant temperature. 
Many of the common spectral features observed in typical SNe~IIP are
 observed in the spectra of all three SNe, 
although the strength of several of the
lines in the red portion of SN~1987A's 
spectrum appear to be larger. This is particularly the case for 
\ion{Na}{i} and \ion{Ba}{ii}  $\lambda$6142. The strong \ion{Ba}{ii} absorption 
line was suggested to be a signature of s-process element 
enhancements in the progenitor of SN~1987A 
(\citealp{williams87}, but see also \citealp{utrobin05}).
 
Our objects exhibit fainter \ion{Ba}{ii} features than SN~1987A (as is also the case for SN~2009E, see \citealp{pastorello11}).
That is likely due to the higher temperature of the photosphere 
(see Fig.~\ref{bolo}, middle panel). 
This can be verified noting that objects showing stronger \ion{Ba}{ii} lines have also lower
temperatures (SN~1997D, \citealp{turatto98}, and similar events, and the previously mentioned 
SNe~2009E and 1987A), whereas high ejecta temperature
objects only show hints of such lines (or nothing at all; see e.g. SN~2009kf, \citealp{botticella10}).
However, intrinsic differences in the composition can not be excluded.
\citet{williams87} also considered \ion{Sr}{ii} lines as a signature of 
s-process element enhancements. Both our objects exhibit a strong
absorption line at 4077~\AA\ which might be associated with this ion. 
In addition, the overall line locations in SN~2006au
are shifted to the blue compared to the other objects, which is 
expected from what was shown in Fig.~\ref{velocity}.

\section{Analysis and discussion}
\label{sec:discussion}

The remarkable similarity between the light curve shapes of 
SNe~2006V and 2006au compared to SN~1987A
suggests that these objects are also mainly powered by the
radioactive decay of $\Ni$. Moreover, early luminosities of SN~1987A-like events are fainter than those of 
normal SNe IIP (see e.g. \citealp{bersten09}).  
This is consistent with a scenario 
involving a relatively compact progenitor star
which spends the majority of its explosion energy on adiabatic expansion. 

In order to derive the physical parameters of the progenitors, a reasonably accurate  
estimation of the explosion dates and the computation of the bolometric light curves are required. In the following two
subsections we discuss the methods and the results concerning these two aspects.
  
\subsection{Explosion dates from the expanding photospheric method}
\label{date}
Using the light curve of SN~1987A as a template, \citet{pastorello05} assumed 
that the peak for SN~1998A was coincident in phase with the peak of 
SN~1987A. This was also consistent with the constraints from the
last non-detection epoch close to discovery.

The discovery date of \V\ is $t_{disc}(06V)=2453771.2$~$JD$, about 45 days from the last non-detection,
$t_{nd}(06V)=2453726.4$~$JD$ \citep{chen06}. For \au\ this interval is even longer, since
 the discovery, $t_{disc}(06au)=2453801.7$~$JD$, occured more than one year since the previous non-detection, $t_{nd}(06au)=2453265.65$~$JD$ \citep{trondal06}. 

Therefore we explore an alternative approach for the determination of the explosion
date. Having a set of 9 spectra for each SN, we used 
the expanding photospheric method (EPM), 
as presented by \citet{jones09}. 
This method is generally used to find the distance of SNe IIP. 
In our case we have
  already an independent measure of the distance ($D$) and we can instead 
use the EPM to constrain the explosion 
date ($t_0$), by using the following procedure: 
 
(i) We checked and,  if necessary, adjusted the flux-calibration of our observed spectra by 
matching synthetic photometry 
to the broad-band photometry.
 
(ii) We then computed synthetic magnitudes 
in the Johnson-Kron-Cousin $VI$ filters
from each redshift-corrected spectrum. 
Here only spectra  obtained prior to $B_{\rm max}$
were considered 
since
the EPM method is most reliable 
at these phases.

(iii)  A black-body spectrum was fit to 
the $VI$ synthetic magnitudes in order to estimate the temperature, $T$, and 
thereby the photospheric radius divided by the SN distance, $R/D$, 
multiplied by the dilution factor, $\zeta$.
   In the SN ejecta the thermalization surface emits radiation, which is diluted by scattering
    before reaching the observer. The dilution is mainly affected by the temperature, as shown in \citet{eastman96} and \citet{dessart05}.

(iv) The dilution factor is estimated according to 
\citet[][see their Table~2]{jones09}  using the 
 expression given for temperatures computed from $V$ and $I$-band  synthetic magnitudes 
and redshift $z=0$. In doing so the expression based on the atmosphere modeling
  of \citet{dessart05} was adopted. The small dependence of the dilution factor on density is ignored.
  
 (v) For each spectrum we considered the velocity, $v$, obtained from the blueshift of the  \ion{Fe}{ii} $\lambda$5169 absorption,
  which is a reasonable proxy for  the photosphere velocity \citep[e.g.][see their Fig. 14]{dessart05}.
 
(vi) Using ${R\zeta}/{D}={v}/{D}(t-t_0)$, 
  an explosion date was then computed for each spectrum and the final value $t_0$ was obtained through their weighted mean. 
  
  The resulting explosion epochs are 
$t_0(06V)=2453748\pm4$ JD and $t_0(06au)=2453794\pm9$ JD. 
These explosion dates are consistent
 with the constraints provided by the discovery dates
 and by the last non-detection dates.
With the aid of SNID, the first spectrum of \au, taken on JD 2453808.9, was compared to the library of spectra of SN~1987A. 
This exercise suggests this spectrum was most similar to SN~1987A approximately 10.9 days after the explosion.
If we assume the same phase for our spectrum, a slightly later (4 days) explosion date for \au\ is obtained. This 
 is consistent with the confidence interval obtained from the EPM analysis. Adopting the EPM estimate of $t_0$,
  the shock break-out cooling tail is found to cover a period of $\sim10-15$~days, which is similar to that of SN~1993J \citep{richmond94}.
 In the case of \V, a SNID analysis did not provide a clear-cut estimate for the phase of our earliest spectra.

 \subsection{Bolometric light curves}

\begin{figure}
\centering 
\includegraphics[width=6in]{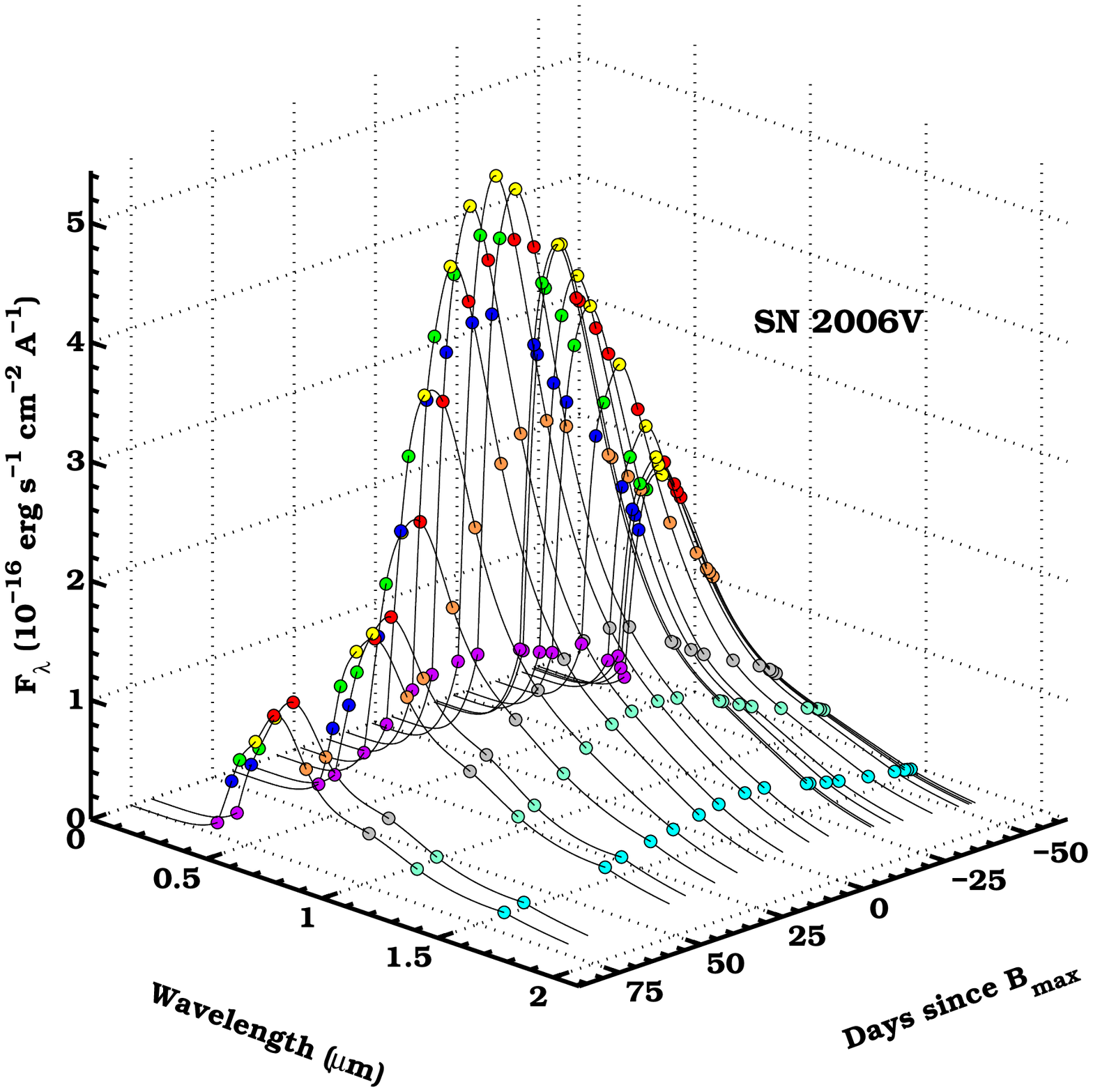}
\caption{\label{SED06V} Spectral energy distributions as a function of time for SN~2006V. The colors correspond to those adopted for each band in Fig.~\ref{lc06V}. The spline fit along with the IR and UV tails are also shown (solid line).}
\end{figure}

\begin{figure}
\centering 
\includegraphics[width=6in]{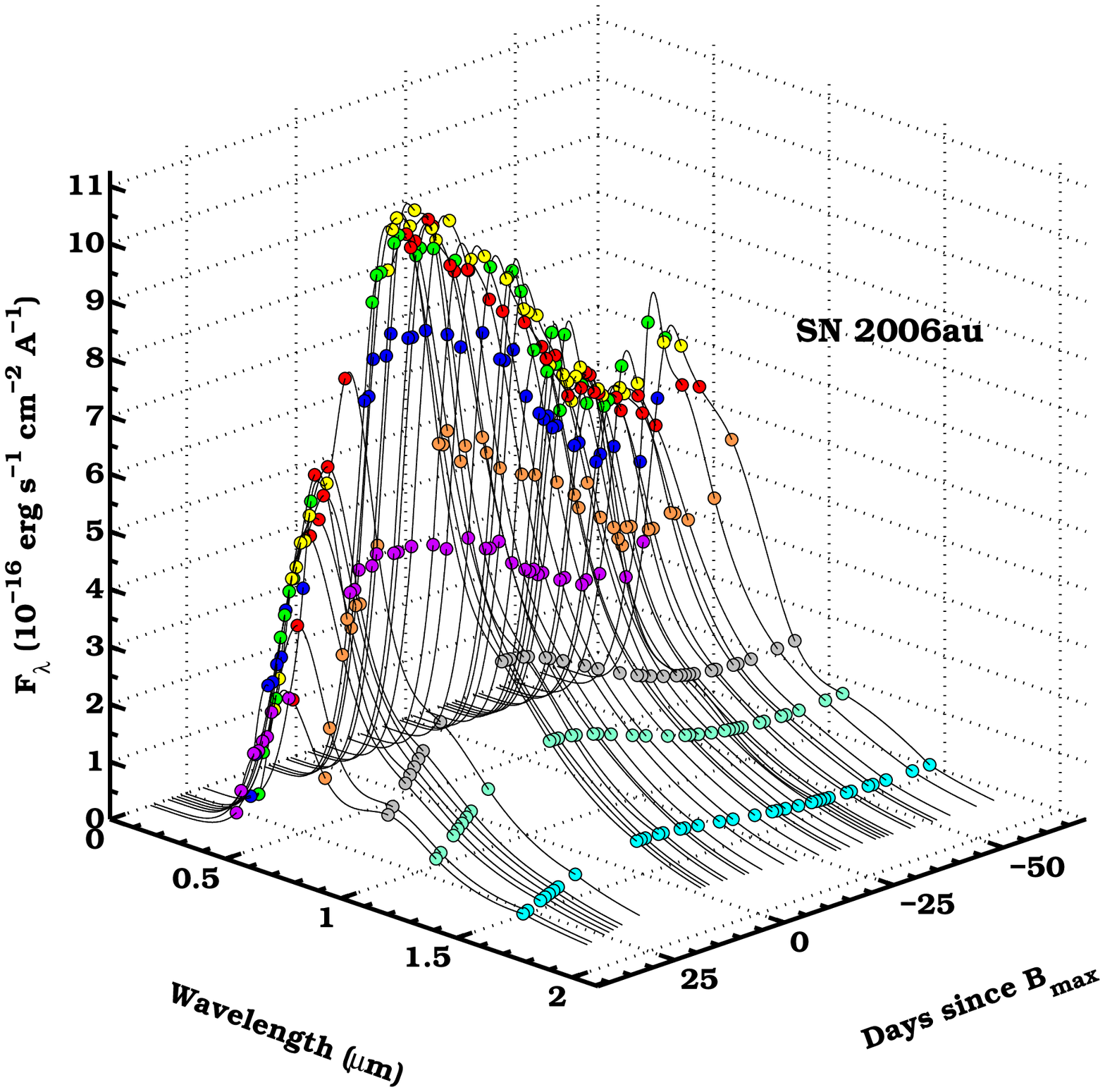}
\caption{\label{SED06au} Spectral energy distributions as a function of time for SN~2006au. The colors correspond to those adopted for each band in Fig.~\ref{lc06au}. The spline fit along with the IR and UV tails are also shown (solid line).}
\end{figure}

The broad wavelength coverage afforded by our multi-band
observations allows us to construct nearly complete 
bolometric light curves of SNe~2006V and 2006au. 
To do so the near-infrared 
light curves were first interpolated to obtained magnitudes at 
the epochs observed in the optical. 
The optical and near-infrared magnitudes were 
converted to flux at the 
effective wavelength of each filter. 
To overcome the lack of $u$-band photometry, magnitudes were 
extrapolated assuming,
$u=B-<B-u>$, with the average $<B-u>$ measured during the rising phase.
In doing so, we assumed a $\pm$0.6~mag error 
in each extrapolated point.
Next a cubic spline was fit to the SEDs at each observed epoch, and this 
is integrated over wavelength. In addition, 
a Rayleigh-Jeans (RJ) tail was included to the overall SED 
to account for flux redwards of the $H$ band, along with a Wien tail to cover the UV flux.
The resulting SEDs are plotted in 
Figs.~\ref{SED06V} and \ref{SED06au}, along with 
the spline fits and the UV and IR tails.
We elected to fit a spline rather than a blackbody to avoid 
over-estimating the flux in the $u$
and $B$ bands \citep{terndrup88}.
As shown by \citet{hamuy88}, 
the inclusion of a  Rayleigh-Jeans tail is a good approximation for 
the case of SN~1987A. 

Plotted in Fig.~\ref{bolo} (top panel) are the bolometric light curves of SNe~2006V, 2006au and 1987A.
Our bolometric light curve of SN~1987A differs slightly from the one in
\citet{hamuy88} because we used a higher 
extinction estimate, $E(B-V)_{tot}=0.175$~mag \citep{woosley87}. 
As indicated by these light curves, SNe~2006V and 2006au 
are more luminous than SN~1987A, reaching peak luminosities of  
log$_{10}(L)\sim$~42.25~erg~s$^{-1}$  and log$_{10}(L)\sim$~42.2~erg~s$^{-1}$, respectively. The peak luminosity of SN~1987A is 
 $\sim~50\%$ fainter, with log$_{10}(L)\sim$~41.90~erg~s$^{-1}$.
 
\subsection{Physical parameters of the progenitors}
  
\begin{figure}
\centering 
\includegraphics[width=6in]{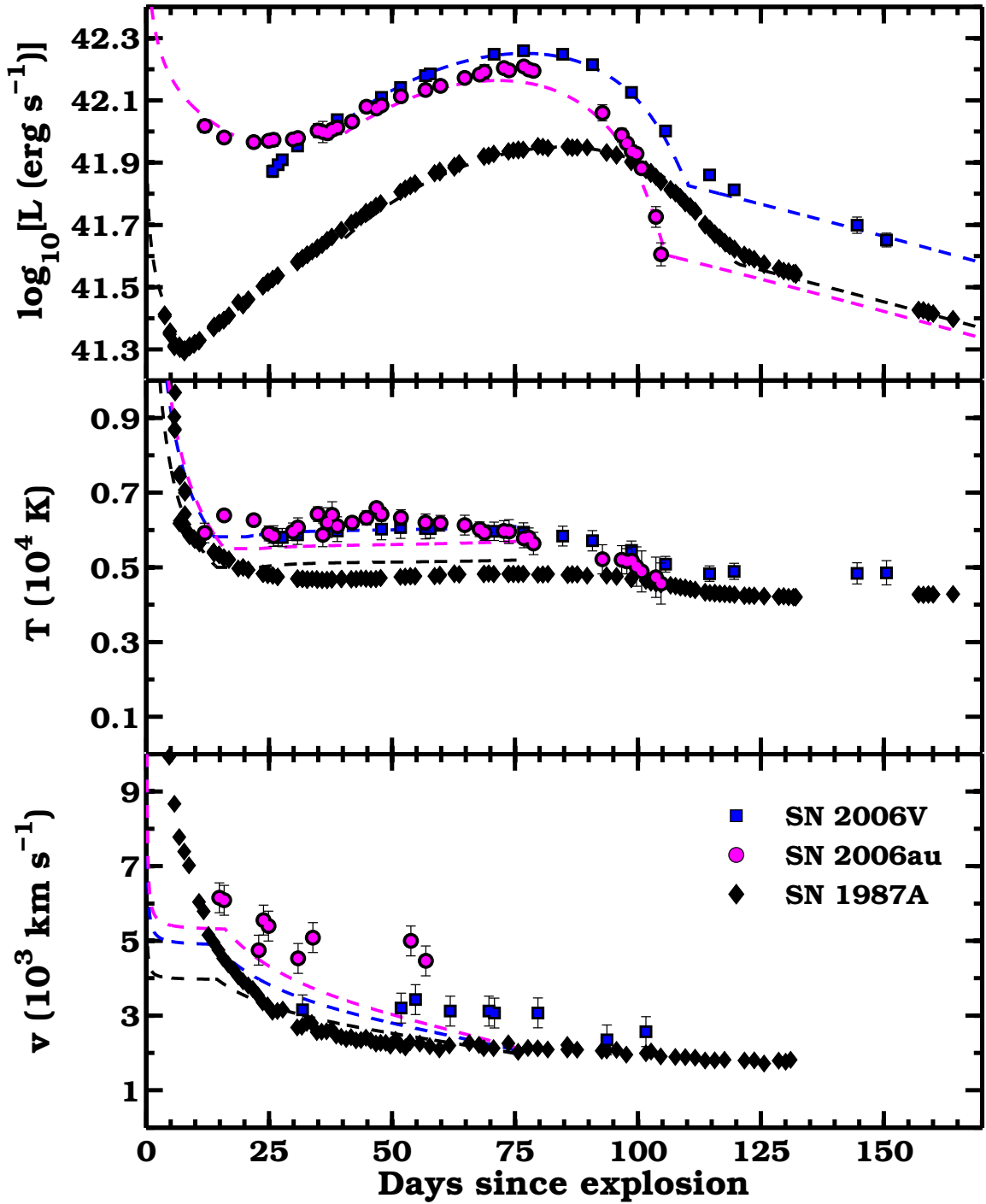}
\caption{\label{bolo}  {\em Top panel:} The bolometric light curves of SNe 
2006V, 2006au and~1987A. Filled symbols show the luminosity resulting from the integration of the SEDs. The fit of the semi-analytic model from \citet{impop92} is represented with dashed lines. {\em Middle panel:} Temperature evolution 
as obtained through black-body fits to the 
 SED for each photometric epoch. {\em Bottom panel:} Photospheric velocity evolution for SNe 2006V, 2006au and~1987A.
For clarity, the errors on temperature and velocity 
for SN~1987A have been omitted in the plot.
These are of the same order as the errors for  SNe 
2006V and 2006au.}
\end{figure}

 With the estimates of $t_0$ from the EPM analysis and the computed bolometric light curves in hand, we next proceed
 to determine the physical parameters of the progenitors of both our SNe. Initial
  rough estimates of the ejected mass ($M_{ej}$) and of the explosion energy ($E$) can be obtained if we
 simply scale these parameters with respect to what has been 
determined for SN~1987A; $E_{87A}=1.1\times10^{51}$~erg  and
$M_{ej}(87A)_{env}=14$~$\Msun$ \citep{blinnikov00}.
In order to do that, we can use the relation $t_d\propto(\kappa M_{ej}/v)^{1/2}$ from \citet{arnett79}, where
 $t_d$ is the diffusion time, $\kappa$ is the mean opacity, and $v$ is a measure of the expansion velocity.
 The relation between the diffusion time of our objects and SN~1987A is $t_d(06V)=0.90~t_d(87A)$ and 
$t_d(06au)=0.85~t_d(87A)$, as measured from the time of the bolometric peaks, see top panel in Fig.~\ref{bolo}. 
If we now assume the same mean opacity for each SN, and use an 
average ratio between the expansion velocities as measured
from the \ion{Fe}{ii} at $\lambda$5169, ($\sim$~1.4 for \V\ and $\sim$~1.7 for \au\, see bottom panel in Fig.~\ref{bolo}),
 the ejecta mass for both objects is found to be 
$M_{ej}\sim20$~$\Msun$ with corresponding ($E\propto M_{ej}v^2$) 
kinetic energies of a few foe. These simple estimates give a first hint on the nature of the progenitors.

From the bolometric light curve one can also constrain 
the amount of \Ni\ synthesized in the explosion. 
The daughter decay product of \Ni\ is 
$^{56}$Co, and it is the decay of $^{56}$Co to $^{56}$Fe 
that powers the late-time light curve.
Unfortunately our photometric coverage of SN~2006au 
does not extend beyond $\sim$~105 
days after the explosion,  so only an upper
limit on the \Ni\ mass can be estimated. 
This is done by assuming the last photometric
epoch belongs to the linear decay phase, 
where $L\propto M_{\Ni}e^{-t/\tau_{}}$. 
We thus obtain 
$M_{\Ni}(06au)\leq 0.073~\Msun$. 
In contrast, the photometric coverage of SN~2006V (up to 150 days after the explosion) is sufficient to directly measure
the \Ni\ mass from the linear decay phase. 
In doing so, we compute $M_{\Ni}(06V)=0.127\pm 0.010~\Msun$. 

Following the plateau luminosity relation presented by \citet{popov93}, 
$L\propto E^{5/6}M_{ej}^{-1/2}R^{2/3}$, 
and scaling with the radius
of SN~1987A, $R(87A)=3\times 10^{12}$~cm $=43$~$\Rsun$ \citep{woosley88}, 
 a rough estimate of the progenitor radius
of SNe 2006V and 2006au can be inferred. We note that this approach has been 
followed by \citet{kleiser11} for estimating the progenitor radius of SN~2000cb, although
\citet{popov93} developed the plateau luminosity relation for SNe whose emission was nonradioactive. 
Such a simple scaling implies radii of $\lesssim 50~\Rsun$ for both objects, which clearly 
suggests a very compact progenitor for our objects.

To confirm these estimates of the progenitor and explosion parameters, we turn to 
the semi-analytic model of \citet{impop92}. 
 This model includes cooling and recombination, and has been shown to 
 provide a good fit to the bolometric light curve of SN~1987A \citep{impop92}.
 As this model is not applicable to the earliest phases, it is fit only  to the bolometric light curves
of SNe 1987A, 2006V and 2006au 
at epochs after 40 days past explosion.
Adopting a mean opacity $\kappa=0.34$~cm$^2$g$^{-1}$, 
and the same exponential distribution of \Nif\
 as used by  \citet{impop92} to fit the bolometric light curve of SN~1987A, 
 $M_{ej}$, $E$, $M_{\Ni}$ and $R$ are estimated  for all three objects.

In the case of \au\ we assume a $M_{\Ni}$ that is equal to the upper limit
estimated from the last photometric epoch. On the other hand, for this SN we can provide additional constraints, in
 particular on the progenitor radius, by using the initial dip in the light curve. 
  This was done for SN~1987A, whose
 light curves also contain a shock break-out cooling tail. 
\citet{chevalier92} provides
an analytical expression for the luminosity of the early-time light curve 
of SN~1987A and similar SNe, in terms of $E$, $M_{ej}$ and $R$.
The luminosity function given by \citet{chevalier92} is
$L=3.08\times10^{43}E_{51}^{0.91}M_{16\Msun}^{-0.40}(F1/1.35)^{-0.17}R_{30\Rsun}t^{-0.34}$
erg~s$^{-1}$. 
$F1$ is the factor by which each gas element increases 
in velocity from $t=0$ to very late times. Following \citet{chevalier92}
we adopt $F1=1.35$. 
As this expression overestimates the luminosity of SN~1987A by a 
factor of 2, we scale it by this factor in order to 
fit the early epochs for both SNe~1987A and 2006au.

The best simultaneous fits to the bolometric light curves 
with the \citet{impop92} model and the \citet{chevalier92} analytic expression 
are shown in Fig.~\ref{bolo} (top panel) as dashed lines. 
The  \citet{impop92} model allows us to also constrain the 
ionization temperature, $T_{ion}$, which strongly affects the light curve 
shape, by fitting the effective temperature before $B_{max}$ 
(dashed lines in the middle panel of Fig.~\ref{bolo}). During the 
recombination phase the photospheric velocity estimated from \ion{Fe}{ii}~$\lambda$5169 has also been fit 
in order to better constrain the energy and mass (dashed lines in the bottom panel of Fig.~\ref{bolo}). 

Our estimates for \V\ are as follows: 
$M_{ej}(06V)=17.0$~$\Msun$, 
$E(06V)=2.4\times10^{51}$~erg, 
$M_{\Ni}(06V)= 0.127$~$\Msun$ and 
$R(06V)=75$~$\Rsun$. 

The model for \au\ gives: 
$M_{ej}(06au)=19.3$~$\Msun$, 
$E(06au)=3.2\times10^{51}$~erg, 
$M_{\Ni}(06au)= 0.073$~$\Msun$ and 
$R(06au)=90$~$\Rsun$. 

The parameters for SN~1987A are: 
$M_{ej}(87A)=11.8$~$\Msun$, 
$E(87A)=1.1\times10^{51}$~erg, 
$M_{\Ni}(87A)= 0.078$~$\Msun$ and 
$R(87A)=33$~$\Rsun$.
The latter is 
in reasonable agreement with the values inferred 
from hydrodynamical simulations \citep{blinnikov00}.

Clearly the adopted semi-analytic model relies on significant simplifications and therefore one can not expect an exact fit to the data.  
However, it does provide a set of reasonable physical parameters.
The mass and energy estimates of our objects are somewhat lower than 
those obtained from the simple scaling relations, and at the same time, are larger than those obtained for SN~1987A. 
The progenitor radii estimates from the semi-analytic models are larger than those estimated from the simple scalings, and
somewhat larger than what is computed for  the progenitor of SN~1987A.
Nevertheless, our estimates on the radii suggests that the progenitors of SNe 2006V and 2006au were compact stars.
 Note that if the same elapsed time between epochs of explosion and maximum for SNe~2006V and 2006au had been assumed
as for SN~1987A,  larger estimates of mass and energy would have been obtained, especially for \au\ ($\sim30$~$\Msun$). 
 The $M_{\Ni}$ estimate would also have been slightly enhanced, while the radius estimate would not change significantly. 
 Even in this case the compact star scenario would be favoured.

\subsection{1987A-like and BSG Supernovae}\label{subsec:87alike}

SNe~2006V and 2006au are important additions to the 
small family of 1987A-like supernovae. The observational properties
of these SNe are summarized by \citet{pastorello11}, and we note that
our SNe are among the best sampled in that collection. 
Among the five best studied
1987A-like objects (SNe~1998A, 2000cb, 2006V, 2006au and 2009E), our two SNe display the intrinsically brightest light curves
 \cite[see][ their Table 5]{pastorello11}. 
Table~\ref{param} contains a summary 
of the derived physical parameters of these five objects.  
Here one  can see that this family of SNe all appear to have ejecta 
masses of about $20~\Msun$. The amount of ejected radioactive nickel is also
consistently $\sim0.1~\Msun$. 
With the exception of SN~2009E, the members of this group
are energetic versions of 
SN~1987A, as elaborated in the case of  SN~2000cb by \citet{utrobin11}. 
Finally, the radii of these stars are consistently found to be relatively 
small ($\lesssim100~\Rsun$). This provides the strongest argument
for their  progenitors to be BSGs.

\begin{figure}
\centering 
\includegraphics[width=6in]{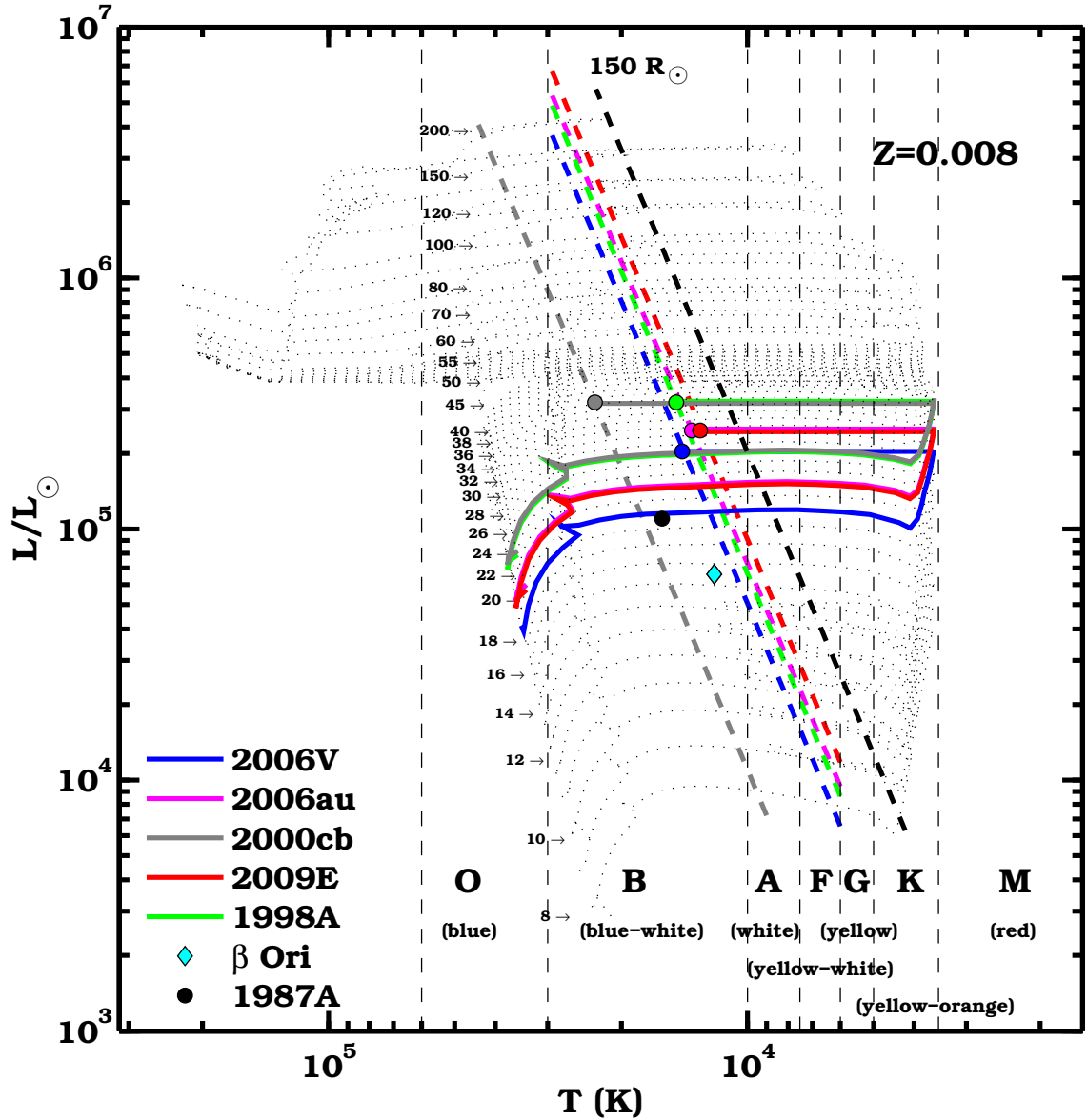}
\caption{\label{HR} Hertzsprung-Russel diagram for 1987A-like SN progenitors. The evolutionary path for each object (colored solid lines) has been computed with the STARS code \citep{eldridge04},
assuming the progenitor mass from the modeling (ejecta mass plus 2~$\Msun$ forming the central compact object) and LMC metallicity ($Z=0.008$, \citealp{glatt10}). The paths for masses ranging between 8 and 200~$\Msun$ are also shown (dotted lines). The position of each progenitor at the moment of the explosion (colored circles) has been obtained by assuming the progenitor to be a black body emitting with the luminosity of the last epoch
of its evolutionary path and having a radius (colored dashed lines) equal to our model estimate. All the 1987A-like SN progenitors belong to the blue-white (B) spectral type as the progenitor of SN~1987A \citep{woosley88}. 
The $150~\Rsun$ line is shown in the plot for comparison, as well as the position of a well-know BSG, $\beta$~Orionis \citep{stewart09}.}
\end{figure}

We note that an estimate of the radius does not univocally determine 
the color of the progenitor star. Indeed some yellow and red supergiants may also have radii of $\approx100~\Rsun$.
A luminosity estimate of the progenitor is also required to infer its temperature, and
 this can be provided by the final epoch of the evolutionary track within the
  Hertzsprung-Russel (HR) diagram as computed with the STARS code \citep{eldridge04}. The
   path is computed according to the mass estimate of each progenitor, roughly given by the
    ejecta mass from our model, and adding 2~$\Msun$ to account for the central compact object. The computed Zero Age Main Sequence (ZAMS) masses belong to the range 19-24~$\Msun$. 
  The HR path for each 1987A-like SN in Table~\ref{param} is shown in Fig.~\ref{HR}. 
This shows that the progenitors of these SN~1987A-like events clearly belong to the B (blue-white) spectral type, as was also the case for the directly
  detected progenitor of SN~1987A \citep{woosley88}. This spectral type is characterized 
  by temperatures between 10000 and 30000~K, and we do find all the progenitors within this range. 
 Note, however, that the evolutionary tracks from the STARS code do not
  actually contain $\sim20~\Msun$ stars which end their lives as BSGs, 
since in their final state they are all positioned in the RSG part of the diagram
 (in Fig.~\ref{HR} the inversions of the evolutionary paths towards the BSG state have been drawn to be consistent with the computed radii).
The physical reason why $\sim20~\Msun$ stars do explode as BSG was discussed
already for SN 1987A \citep{arnett89,pod92}, but it is still not settled if this
is due to metallicity or binary evolution. 

In any case, the characteristic small progenitor radii in conjunction with 
the relatively high ejecta masses of these 1987A-like SNe allow us to reject a 
red supergiant origin.
Even the yellow supergiant progenitors that have recently been 
suggested \cite[see][ their Figure 4]{maund11} favour
a position in the HR diagram compatible with larger radii ($\sim300~\Rsun$).

The final explosion of a BSG can apparently proceed in different ways, 
as can be seen from the diversity that is present in Table~\ref{param}. 
It would be of interest to enlarge this sample, and 
to model them all in a consistent way. Such an effort is underway \cite[see][]{pastorello11}. It is also clear
 that these events are intrinsically quite rare, and this is likely the reason 
 why these massive BSG progenitors have not yet shown up in direct detections \citep{smartt09}.

\section{Conclusions}\label{sec:conclusions}

We have presented extensive observations of two SNe
that have been shown to exhibit similarities to SN~1987A.
Just as SN~1987A, both objects show a slow rise to the peak, with SN~2006V
reaching $B_{\rm{max}}$  76 days past explosion  and SN~2006au  after 72 days.
Detailed inspection show, however, that these SNe display faster 
velocities and higher luminosities  compared to SN~1987A. These observational 
parameters also suggest that our objects have higher explosion energies
than SN~1987A.

\V\ shows the highest luminosity, and 
produced the largest amount of \Nif, $M_{\Ni}(06V)= 0.127$~$\Msun$.
For \au\  only an upper limit on the \Ni\ mass can be made, $M_{\Ni}(06au) 
\leqslant  0.073$~$\Msun$.
However, this is comparable to what SN~1987A produced.

A semi-analytic model was applied to the bolometric light 
curves of both SNe 2006V and 2006au that suggests the radii of these
objects are consistent with a BSG progenitor. 
When comparing the small family of objects in the literature 
thought to have a BSG origin, we find that their progenitors
exhibit a wide range in masses, radii, explosion energies, and 
\Nif\ production.

\begin{acknowledgements}

The Oskar Klein Centre is funded by the Swedish Research Council.
The Dark Cosmology Centre is funded by the Danish National Research Foundation. 
This material is based upon work supported by NSF under 
grants AST--0306969, AST--0908886, AST--0607438, and AST--1008343. 

M. Hamuy and J.P. Anderson acknowledge support by CONICYT through
FONDECYT grants 1060808 and 3110142, Centro de Astrofisica FONDAP
15010003, Centro BASAL CATA (PFB-06), and by the Millenium Center for
Supernova Science (P06-045-F, P10-064-F).

We acknowledge  C. Burns, L. Boldt, C. Contreras,  A. Campillay,
 W. Krzeminski, M. Roth,  F. Salgado, B. Madore and N. Suntzeff for their support.

\end{acknowledgements}

\bibliographystyle{aa}

\onecolumn

\clearpage
\begin{deluxetable} {lllccccccccc}
\rotate
\tabletypesize{\tiny}
\tablecolumns{12}
\tablewidth{0pt}
\tablecaption{Optical and near-infrared photometry of the local sequences in the standard system.\label{tabstars}}
\tablehead{
\colhead{} &
\colhead{}  &
\colhead{}  &
\colhead{$u$}   &
\colhead{$g$}   &
\colhead{$r$}   &
\colhead{$i$}   &
\colhead{$B$}   &
\colhead{$V$}  &
\colhead{$Y$}   &
\colhead{$J$}   &
\colhead{$H$}  \\
\colhead{STAR} &
\colhead{$\alpha(2000)$}  &
\colhead{$\delta(2000)$}  &
\colhead{(mag)}   &
\colhead{(mag)}   &
\colhead{(mag)}   &
\colhead{(mag)}   &
\colhead{(mag)}   &
\colhead{(mag)}  &
\colhead{(mag)}   &
\colhead{(mag)}   &
\colhead{(mag)}}
\startdata
\multicolumn{12}{c}{\bf SN 2006V}\\
01	& 11:31:37.71	& $-$02:18:48.30	& $ \ldots $	& $ \ldots $	& $ \ldots $	& $ \ldots $	& $ \ldots $	& $ \ldots $	&  12.444(006)	&  12.163(006)	&  11.800(006)		\\
02	& 11:31:21.53	& $-$02:20:22.56	&  17.469(072)	&  15.625(011)	&  14.923(011)	&  14.669(011)	&  16.091(009)	&  15.227(010)	& $ \ldots $	& $ \ldots $	& $ \ldots $	 	\\
03	& 11:31:45.09	& $-$02:17:18.60	&  17.312(018)	&  16.154(011)	&  15.723(011)	&  15.570(011)	&  16.494(009)	&  15.900(008)	&  14.977(006)	&  14.715(006)	&  14.379(006)	 	\\
04	& 11:31:20.58	& $-$02:18:32.81	&  17.503(013)	&  16.226(013)	&  15.789(011)	&  15.643(011)	&  16.588(009)	&  15.963(008)	& $ \ldots $	& $ \ldots $	& $ \ldots $	 	\\
05	& 11:31:38.33	& $-$02:21:37.42	&  18.270(026)	&  16.697(011)	&  16.089(011)	&  15.876(013)	&  17.130(011)	&  16.334(010)	&  15.172(009)	&  14.863(008)	&  14.434(007)	 	\\
06	& 11:31:32.71	& $-$02:21:53.30	&  19.600(084)	&  17.251(012)	&  16.221(018)	&  15.854(011)	&  17.819(011)	&  16.686(010)	&  14.980(009)	&  14.583(009)	&  14.014(009)	 	\\
07	& 11:31:41.76	& $-$02:16:13.31	&  19.153(052)	&  17.342(009)	&  16.564(010)	&  16.251(010)	&  17.837(009)	&  16.900(008)	&  15.450(006)	&  15.079(008)	&  14.584(008)	 	\\
08	& 11:31:26.83	& $-$02:19:17.16	&  20.471(180)	&  18.008(009)	&  16.735(011)	&  16.181(009)	&  18.673(012)	&  17.310(008)	&  15.257(079)	&  14.963(147)	&  14.301(130)	 	\\
09	& 11:31:45.74	& $-$02:19:25.57	& $ \ldots $	&  18.437(009)	&  17.035(012)	&  15.927(010)	&  19.252(018)	&  17.638(008)	&  14.585(007)	&  14.108(008)	&  13.582(006)	 	\\
10	& 11:31:32.39	& $-$02:15:16.22	&  19.956(126)	&  18.008(009)	&  17.156(011)	&  16.824(009)	&  18.532(010)	&  17.547(008)	& $ \ldots $	& $ \ldots $	& $ \ldots $	 	\\
11	& 11:31:45.57	& $-$02:16:45.48	&  18.951(055)	&  18.048(012)	&  17.753(009)	&  17.629(013)	&  18.296(014)	&  17.869(008)	&  17.100(017)	&  16.828(019)	&  16.554(034)	 	\\
12	& 11:31:47.18	& $-$02:21:27.35	& $ \ldots $	&  19.166(029)	&  17.843(011)	&  17.176(013)	&  19.931(040)	&  18.446(011)	& $ \ldots $	& $ \ldots $	& $ \ldots $	 	\\
13	& 11:31:28.90	& $-$02:17:16.32	&  20.975(281)	&  19.186(015)	&  17.980(019)	&  17.449(009)	&  19.813(026)	&  18.536(014)	&  16.487(010)	&  16.086(054)	&  15.480(054)	 	\\
14	& 11:31:36.82	& $-$02:19:57.30	& $ \ldots $	&  19.102(023)	&  17.975(015)	&  17.489(009)	&  19.726(041)	&  18.498(009)	&  16.510(016)	&  16.078(009)	&  15.480(016)	 	\\
15	& 11:31:29.08	& $-$02:14:38.58	&  20.088(108)	&  18.906(020)	&  18.367(009)	&  18.169(010)	&  19.284(017)	&  18.601(013)	& $ \ldots $	& $ \ldots $	& $ \ldots $	 	\\
16	& 11:31:36.40	& $-$02:18:16.45	&  19.388(265)	&  18.913(013)	&  18.429(012)	&  18.229(011)	&  19.280(031)	&  18.611(017)	& $ \ldots $	& $ \ldots $	& $ \ldots $	 	\\
17	& 11:31:26.40	& $-$02:20:05.83	& $ \ldots $	&  20.004(024)	&  18.559(027)	&  17.090(017)	&  20.856(075)	&  19.157(023)	& $ \ldots $	& $ \ldots $	& $ \ldots $	 	\\
18	& 11:31:43.91	& $-$02:18:40.94	&  20.305(132)	&  19.117(015)	&  18.576(009)	&  18.365(013)	&  19.514(027)	&  18.775(021)	&  17.652(045)	&  17.326(046)	&  17.029(061)	 	\\
19	& 11:31:37.16	& $-$02:14:50.08	& $ \ldots $	&  19.659(018)	&  18.669(027)	&  18.334(011)	&  20.176(060)	&  19.116(038)	& $ \ldots $	& $ \ldots $	& $ \ldots $	 	\\
20	& 11:31:25.90	& $-$02:20:41.30	&  20.213(116)	&  19.340(014)	&  18.950(012)	&  18.775(026)	&  19.606(032)	&  19.074(015)	& $ \ldots $	& $ \ldots $	& $ \ldots $		\\
21	& 11:31:28.66	& $-$02:19:25.59	&  20.534(180)	&  19.626(031)	&  19.279(020)	&  19.190(023)	&  19.902(030)	&  19.407(019)	& $ \ldots $	& $ \ldots $	& $ \ldots $	 	\\
22	& 11:31:42.41	& $-$02:22:43.86	& $ \ldots $	& $ \ldots $	& $ \ldots $	& $ \ldots $	& $ \ldots $	& $ \ldots $	&  14.232(009)	&  13.925(009)	&  13.529(009)	 	\\
23	& 11:31:49.73	& $-$02:14:25.94	& $ \ldots $	& $ \ldots $	& $ \ldots $	& $ \ldots $	& $ \ldots $	& $ \ldots $	&  14.747(073)	&  14.604(062)	&  14.445(054)	 	\\
24	& 11:31:54.45	& $-$02:22:18.01	& $ \ldots $	& $ \ldots $	& $ \ldots $	& $ \ldots $	& $ \ldots $	& $ \ldots $	&  15.653(016)	&  15.177(009)	&  14.676(009)	 	\\
25	& 11:31:55.23	& $-$02:17:25.15	& $ \ldots $	& $ \ldots $	& $ \ldots $	& $ \ldots $	& $ \ldots $	& $ \ldots $	&  15.629(007)	&  15.200(006)	&  14.647(006)	 	\\
26	& 11:31:30.39	& $-$02:14:03.77	& $ \ldots $	& $ \ldots $	& $ \ldots $	& $ \ldots $	& $ \ldots $	& $ \ldots $	&  15.998(009)	&  15.494(011)	&  14.994(017)	 	\\
27	& 11:31:46.96	& $-$02:22:02.93	& $ \ldots $	& $ \ldots $	& $ \ldots $	& $ \ldots $	& $ \ldots $	& $ \ldots $	&  16.188(013)	&  15.690(013)	&  15.049(019)	 	\\
28	& 11:31:29.32	& $-$02:18:19.33	& $ \ldots $	& $ \ldots $	& $ \ldots $	& $ \ldots $	& $ \ldots $	& $ \ldots $	&  16.507(043)	&  16.020(074)	&  15.513(075)	 	\\
29	& 11:31:49.54	& $-$02:23:03.84	& $ \ldots $	& $ \ldots $	& $ \ldots $	& $ \ldots $	& $ \ldots $	& $ \ldots $	&  16.775(019)	&  16.229(021)	&  15.691(040)	 	\\
30	& 11:31:49.62	& $-$02:16:20.46	& $ \ldots $	& $ \ldots $	& $ \ldots $	& $ \ldots $	& $ \ldots $	& $ \ldots $	&  16.948(014)	&  16.426(015)	&  15.834(021)		\\
31	& 11:31:27.00	& $-$02:19:43.97	& $ \ldots $	& $ \ldots $	& $ \ldots $	& $ \ldots $	& $ \ldots $	& $ \ldots $	&  17.058(015)	&  16.625(081)	&  15.985(025)	 	\\
32	& 11:31:27.36	& $-$02:22:40.51	& $ \ldots $	& $ \ldots $	& $ \ldots $	& $ \ldots $	& $ \ldots $	& $ \ldots $	&  17.038(031)	&  16.585(021)	&  15.955(037)	 	\\
33	& 11:31:57.32	& $-$02:21:26.93	& $ \ldots $	& $ \ldots $	& $ \ldots $	& $ \ldots $	& $ \ldots $	& $ \ldots $	&  17.339(033)	&  16.959(033)	&  16.270(032)		\\
34	& 11:31:34.96	& $-$02:18:01.44	& $ \ldots $	& $ \ldots $	& $ \ldots $	& $ \ldots $	& $ \ldots $	& $ \ldots $	&  17.382(014)	&  16.964(015)	&  16.408(049)	 	\\
35	& 11:31:55.07	& $-$02:15:07.24	& $ \ldots $	& $ \ldots $	& $ \ldots $	& $ \ldots $	& $ \ldots $	& $ \ldots $	&  17.237(024)	&  16.770(024)	&  16.189(039)	    \\
36	& 11:31:35.43	& $-$02:18:29.34	& $ \ldots $	& $ \ldots $	& $ \ldots $	& $ \ldots $	& $ \ldots $	& $ \ldots $	&  17.526(027)	&  17.067(083)	&  16.451(041)	 	\\
37	& 11:31:31.01	& $-$02:23:37.25	& $ \ldots $	& $ \ldots $	& $ \ldots $	& $ \ldots $	& $ \ldots $	& $ \ldots $	&  17.390(030)	&  16.993(040)	&  16.377(060)	 	\\
38	& 11:31:31.89	& $-$02:14:27.20	& $ \ldots $	& $ \ldots $	& $ \ldots $	& $ \ldots $	& $ \ldots $	& $ \ldots $	&  17.884(042)	&  17.501(059)	&  16.945(065)	 	\\
39	& 11:31:51.38	& $-$02:18:15.59	& $ \ldots $	& $ \ldots $	& $ \ldots $	& $ \ldots $	& $ \ldots $	& $ \ldots $	&  15.977(010)	&  15.492(007)	&  14.990(008)	 	\\
40	& 11:31:41.17	& $-$02:17:06.90	& $ \ldots $	& $ \ldots $	& $ \ldots $	& $ \ldots $	& $ \ldots $	& $ \ldots $	&  18.347(054)	&  17.917(050)	&  17.474(094)	 	\\

\multicolumn{12}{c}{\bf SN 2006au}\\
01	& 17:57:01.27	& $+$12:12:17.97	& $ \ldots $	& $ \ldots $	& $ \ldots $	& $ \ldots $	& $ \ldots $	& $ \ldots $	& $ \ldots $	& $ \ldots $	& $ \ldots $	 	\\
02	& 17:57:26.02	& $+$12:09:44.36	& $ \ldots $	& $ \ldots $	& $ \ldots $	& $ \ldots $	& $ \ldots $	& $ \ldots $	&  12.676(006)	&  12.432(006)	&  12.153(009)	 	\\
03	& 17:57:25.34	& $+$12:10:46.35	& $ \ldots $	& $ \ldots $	& $ \ldots $	& $ \ldots $	&  15.090(012)	& $ \ldots $	& $ \ldots $	& $ \ldots $	& $ \ldots $	 	\\
04	& 17:57:24.92	& $+$12:07:12.78	& $ \ldots $	& $ \ldots $	& $ \ldots $	& $ \ldots $	& $ \ldots $	& $ \ldots $	& $ \ldots $	& $ \ldots $	& $ \ldots $	 	\\
05	& 17:57:23.47	& $+$12:13:48.50	& $ \ldots $	& $ \ldots $	& $ \ldots $	& $ \ldots $	& $ \ldots $	& $ \ldots $	& $ \ldots $	& $ \ldots $	& $ \ldots $	 	\\
06	& 17:57:21.28	& $+$12:11:19.12	& $ \ldots $	& $ \ldots $	& $ \ldots $	& $ \ldots $	& $ \ldots $	& $ \ldots $	&  11.928(006)	&  11.703(009)	&  11.416(009)	 	\\
07	& 17:57:21.20	& $+$12:13:47.45	&  16.051(016)	&  14.925(015)	&  14.566(015)	&  14.447(015)	&  15.203(016)	&  14.664(035)	&  13.745(011)	&  13.557(012)	&  13.344(037)	 	\\
08	& 17:57:20.99	& $+$12:08:56.40	&  16.373(018)	& $ \ldots $	& $ \ldots $	& $ \ldots $	&  15.212(009)	&  14.471(014)	&  13.325(006)	&  13.070(006)	&  12.773(009)	 	\\
09	& 17:57:19.13	& $+$12:12:49.98	& $ \ldots $	& $ \ldots $	& $ \ldots $	& $ \ldots $	& $ \ldots $	& $ \ldots $	& $ \ldots $	& $ \ldots $	& $ \ldots $	 	\\
10	& 17:57:14.53	& $+$12:11:56.61	& $ \ldots $	& $ \ldots $	& $ \ldots $	& $ \ldots $	& $ \ldots $	& $ \ldots $	& $ \ldots $	& $ \ldots $	& $ \ldots $	 	\\
11	& 17:57:10.92	& $+$12:14:24.91	& $ \ldots $	& $ \ldots $	& $ \ldots $	& $ \ldots $	& $ \ldots $	& $ \ldots $	& $ \ldots $	& $ \ldots $	& $ \ldots $	 	\\
12	& 17:57:10.72	& $+$12:12:45.99	& $ \ldots $	& $ \ldots $	& $ \ldots $	& $ \ldots $	& $ \ldots $	& $ \ldots $	&  11.684(006)	&  11.384(020)	&  10.956(020)	 	\\
13	& 17:57:10.58	& $+$12:12:10.49	&  17.710(026)	&  15.397(015)	&  14.531(015)	&  14.210(015)	&  15.962(007)	&  14.884(010)	&  13.311(007)	&  12.961(013)	&  12.471(012)	 	\\
14	& 17:57:07.66	& $+$12:10:05.07	&  17.735(026)	& $ \ldots $	& $ \ldots $	& $ \ldots $	&  15.698(008)	& $ \ldots $	& $ \ldots $	& $ \ldots $	& $ \ldots $	 	\\
15	& 17:57:07.08	& $+$12:12:04.19	& $ \ldots $	& $ \ldots $	& $ \ldots $	& $ \ldots $	& $ \ldots $	& $ \ldots $	& $ \ldots $	& $ \ldots $	& $ \ldots $	 	\\
16	& 17:57:24.01	& $+$12:09:27.97	&  18.151(035)	&  15.910(009)	&  15.040(009)	&  14.698(013)	&  16.497(007)	&  15.399(008)	& $ \ldots $	& $ \ldots $	& $ \ldots $	 	\\
17	& 17:57:07.13	& $+$12:12:39.82	&  17.220(017)	&  15.637(009)	&  15.060(009)	&  14.838(011)	&  16.075(007)	&  15.286(006)	& $ \ldots $	& $ \ldots $	& $ \ldots $	 	\\
18	& 17:57:10.34	& $+$12:10:10.92	&  19.088(082)	&  16.416(007)	&  15.260(007)	&  14.789(009)	&  17.034(007)	&  15.780(007)	&  13.809(007)	&  13.421(014)	&  12.828(035)	 	\\
19	& 17:57:04.06	& $+$12:10:21.84	&  17.593(023)	&  16.090(007)	&  15.519(007)	&  15.292(007)	&  16.512(007)	&  15.743(006)	& $ \ldots $	& $ \ldots $	& $ \ldots $	 	\\
20	& 17:57:05.77	& $+$12:07:10.60	&  18.015(061)	&  16.248(007)	&  15.532(007)	&  15.223(007)	&  16.723(008)	&  15.818(007)	& $ \ldots $	& $ \ldots $	& $ \ldots $	 	\\
21	& 17:57:05.24	& $+$12:08:46.35	&  18.154(038)	&  16.294(007)	&  15.544(007)	&  15.218(007)	&  16.817(007)	&  15.848(006)	& $ \ldots $	& $ \ldots $	& $ \ldots $	 	\\
22	& 17:57:08.21	& $+$12:08:23.53	&  18.508(049)	&  16.707(007)	&  16.046(007)	&  15.798(007)	&  17.176(009)	&  16.315(006)	& $ \ldots $	& $ \ldots $	& $ \ldots $	 	\\
23	& 17:57:26.13	& $+$12:12:18.72	&  19.005(077)	&  16.967(007)	&  16.136(007)	&  15.781(008)	&  17.508(008)	&  16.476(006)	& $ \ldots $	& $ \ldots $	& $ \ldots $	 	\\
24	& 17:57:18.41	& $+$12:11:17.41	&  19.073(129)	&  17.004(007)	&  16.213(007)	&  15.890(007)	&  17.545(008)	&  16.541(006)	&  15.038(008)	&  14.684(013)	&  14.193(013)	 	\\
25	& 17:57:18.10	& $+$12:07:34.41	&  19.381(114)	&  17.338(008)	&  16.478(007)	&  16.094(008)	&  17.881(012)	&  16.847(008)	& $ \ldots $	& $ \ldots $	& $ \ldots $	 	\\
26	& 17:57:17.76	& $+$12:08:23.45	&  18.382(071)	&  17.006(007)	&  16.482(007)	&  16.277(009)	&  17.407(008)	&  16.693(006)	& $ \ldots $	& $ \ldots $	& $ \ldots $	 	\\
27	& 17:57:04.64	& $+$12:11:26.91	&  18.523(052)	&  17.182(008)	&  16.620(007)	&  16.377(007)	&  17.579(009)	&  16.850(007)	& $ \ldots $	& $ \ldots $	& $ \ldots $	 	\\
28	& 17:57:10.82	& $+$12:08:44.31	&  18.985(078)	&  17.505(007)	&  16.955(007)	&  16.745(011)	&  17.922(009)	&  17.171(006)	&  16.010(006)	&  15.737(007)	&  15.393(022)	 	\\
29	& 17:57:24.06	& $+$12:08:07.25	&  19.253(099)	&  17.641(007)	&  17.075(007)	&  16.850(007)	&  18.080(010)	&  17.291(007)	& $ \ldots $	& $ \ldots $	& $ \ldots $	 	\\
30	& 17:57:39.05	& $+$12:08:46.21	& $ \ldots $	& $ \ldots $	& $ \ldots $	& $ \ldots $	& $ \ldots $	& $ \ldots $	&  15.015(007)	&  14.646(007)	&  14.172(013)	 	\\
31	& 17:57:23.88	& $+$12:11:39.48	& $ \ldots $	& $ \ldots $	& $ \ldots $	& $ \ldots $	& $ \ldots $	& $ \ldots $	&  16.760(014)	&  16.507(013)	&  16.237(047)	 	\\
32	& 17:57:29.01	& $+$12:11:17.38	& $ \ldots $	& $ \ldots $	& $ \ldots $	& $ \ldots $	& $ \ldots $	& $ \ldots $	&  16.880(014)	&  16.589(016)	&  16.261(043)	 	\\
33	& 17:57:34.21	& $+$12:12:06.84	& $ \ldots $	& $ \ldots $	& $ \ldots $	& $ \ldots $	& $ \ldots $	& $ \ldots $	&  16.879(013)	&  16.579(014)	&  16.204(033)	 	\\
34	& 17:57:13.41	& $+$12:07:53.58	& $ \ldots $	& $ \ldots $	& $ \ldots $	& $ \ldots $	& $ \ldots $	& $ \ldots $	&  17.032(025)	&  16.706(041)	&  16.217(070)	 	\\
35	& 17:57:40.05	& $+$12:13:03.83	& $ \ldots $	& $ \ldots $	& $ \ldots $	& $ \ldots $	& $ \ldots $	& $ \ldots $	&  17.056(015)	&  16.747(029)	&  16.345(065)	 	\\
36	& 17:57:10.67	& $+$12:09:20.38	& $ \ldots $	& $ \ldots $	& $ \ldots $	& $ \ldots $	& $ \ldots $	& $ \ldots $	&  17.090(016)	&  16.722(015)	&  16.294(048)	 	\\
37	& 17:57:22.71	& $+$12:09:52.74	& $ \ldots $	& $ \ldots $	& $ \ldots $	& $ \ldots $	& $ \ldots $	& $ \ldots $	&  17.127(025)	&  16.656(020)	&  16.096(038)	 	\\
38	& 17:57:40.25	& $+$12:09:44.53	& $ \ldots $	& $ \ldots $	& $ \ldots $	& $ \ldots $	& $ \ldots $	& $ \ldots $	&  17.099(028)	&  16.746(020)	&  16.270(054)	 	\\
39	& 17:57:14.91	& $+$12:13:28.13	& $ \ldots $	& $ \ldots $	& $ \ldots $	& $ \ldots $	& $ \ldots $	& $ \ldots $	&  17.578(045)	&  17.153(032)	&  16.523(157)	 	\\
40	& 17:57:16.09	& $+$12:12:23.22	& $ \ldots $	& $ \ldots $	& $ \ldots $	& $ \ldots $	& $ \ldots $	& $ \ldots $	&  17.656(039)	&  17.261(024)	&  16.621(038)	 	\\
41	& 17:57:20.72	& $+$12:07:54.01	& $ \ldots $	& $ \ldots $	& $ \ldots $	& $ \ldots $	& $ \ldots $	& $ \ldots $	&  17.803(050)	&  17.314(039)	&  16.852(063)	 	\\
42	& 17:57:14.90	& $+$12:11:26.23	& $ \ldots $	& $ \ldots $	& $ \ldots $	& $ \ldots $	& $ \ldots $	& $ \ldots $	&  14.535(007)	&  14.045(012)	&  13.428(007)	 	\\
43	& 17:57:12.58	& $+$12:11:23.64	& $ \ldots $	& $ \ldots $	& $ \ldots $	& $ \ldots $	& $ \ldots $	& $ \ldots $	&  17.510(024)	&  17.171(024)	&  16.558(031)	 	\\
44	& 17:57:16.80	& $+$12:11:22.09	& $ \ldots $	& $ \ldots $	& $ \ldots $	& $ \ldots $	& $ \ldots $	& $ \ldots $	&  16.658(013)	&  16.329(021)	&  15.807(022)	 	\\
45	& 17:57:10.85	& $+$12:11:15.14	& $ \ldots $	& $ \ldots $	& $ \ldots $	& $ \ldots $	& $ \ldots $	& $ \ldots $	&  16.649(027)	&  16.364(041)	&  15.994(061)	 	\\
46	& 17:57:16.99	& $+$12:11:49.88	& $ \ldots $	& $ \ldots $	& $ \ldots $	& $ \ldots $	& $ \ldots $	& $ \ldots $	&  17.242(019)	&  16.910(016)	&  16.380(031)	 	\\
47	& 17:57:12.86	& $+$12:09:42.55	& $ \ldots $	& $ \ldots $	& $ \ldots $	& $ \ldots $	& $ \ldots $	& $ \ldots $	&  16.118(014)	&  15.815(016)	&  15.401(023)	 	\\
48	& 17:57:18.86	& $+$12:12:19.19	& $ \ldots $	& $ \ldots $	& $ \ldots $	& $ \ldots $	& $ \ldots $	& $ \ldots $	&  16.814(010)	&  16.563(018)	&  16.178(029)	 	\\
49	& 17:57:17.80	& $+$12:10:20.24	& $ \ldots $	& $ \ldots $	& $ \ldots $	& $ \ldots $	& $ \ldots $	& $ \ldots $	&  16.227(010)	&  15.801(013)	&  15.130(015)	 	\\
50	& 17:57:22.52	& $+$12:10:57.65	& $ \ldots $	& $ \ldots $	& $ \ldots $	& $ \ldots $	& $ \ldots $	& $ \ldots $	&  16.196(009)	&  15.736(007)	&  15.211(014)	 	\\

\enddata
\tablecomments{Uncertainties given in parentheses in thousandths of a
  magnitude correspond to an rms of the magnitudes obtained on
  photometric nights.}

\end{deluxetable}

\clearpage
\begin{deluxetable}{lclcclcc}
\tabletypesize{\scriptsize}
\tablewidth{0pt}
\tablecolumns{8}
\tablecaption{Spectroscopic observations of SN~2006V and SN~2006au.\label{tabspectra}}
\tablehead{
\colhead{Date} &
\colhead{Julian Date} &
\colhead{Epoch\tablenotemark{a}} &
\colhead{Telescope} &
\colhead{Instrument} &
\colhead{Range} &
\colhead{Resolution} &
\colhead{Integration} \\ 
\colhead{(UT)} &
\colhead{JD - 2,453,000} &
\colhead{(days)} &
\colhead{} &
\colhead{} &
\colhead{(\AA)} &
\colhead{(FWHM \AA)} &
\colhead{(sec)}}
\startdata
\multicolumn{8}{c}{\bf SN~2006V}\\
 2006 Feb. 13 & 779.8   &  $-$43.9    & NTT     & EMMI   & 4000 -- 10200  & 9   &300 \\ 
 2006 Mar. 05 & 799.8   &  $-$23.8    & Du~Pont & WFCCD  & 3800 -- 9235   & 8   &900 \\ 
 2006 Mar. 08 & 802.8   &  $-$20.9    & Du~Pont & WFCCD  & 3800 -- 9235   & 8   &900 \\
 2006 Mar. 15 & 809.8   &  $-$13.8    & Clay    & LDSS   & 3785 -- 9969   & 2-4\tablenotemark{b} &600 \\
 2006 Mar. 23 & 817.7   &  $-$6.0     & Du~Pont & WFCCD  & 3800 -- 9235   & 8   &900 \\
 2006 Mar. 24 & 818.8   &  $-$4.9     & Du~Pont & WFCCD  & 3800 -- 9235   & 8   &900 \\
 2006 Apr. 02 & 827.7   &  $+$4.0     & Du~Pont & WFCCD  & 3800 -- 9235   & 8   &900 \\
 2006 Apr. 16 & 841.7   &  $+$18.0    & Baade   & IMACS  & 3842 -- 9692   & 4   &900 \\
 2006 Apr. 24 & 849.6   &  $+$25.9    & Du~Pont & WFCCD  & 3800 -- 9235   & 8   &900 \\
\hline
\multicolumn{8}{c}{\bf SN~2006au}\\
2006  Mar. 14 & 808.9  &  $-$56.6 & Clay    & LDSS   & 3785 -- 9969   & 2-4\tablenotemark{b}   & 600   \\
2006  Mar. 15 & 809.9  &  $-$55.6 & NTT     & EMMI   & 4000 -- 10200  & 8   & 300   \\
2006  Mar. 22 & 816.9  &  $-$48.6 & Du~Pont & WFCCD  & 3800 -- 9235   & 8   & 600   \\
2006  Mar. 23 & 817.9  &  $-$47.6 & Du~Pont & WFCCD  & 3800 -- 9235   & 8   & 900   \\
2006  Mar. 24 & 818.9  &  $-$46.6 & Du~Pont & WFCCD  & 3800 -- 9235   & 8   & 600   \\
2006  Mar. 30 & 824.9  &  $-$40.6 & Du~Pont & WFCCD  & 3800 -- 9235   & 8   & 900   \\
2006  Apr. 02 & 827.9  &  $-$37.6 & Du~Pont & WFCCD  & 3800 -- 9235   & 8   & 600   \\
2006  Apr. 22 & 847.8  &  $-$17.7 & Du~Pont & WFCCD  & 3800 -- 9235   & 8   & 600   \\
2006  Apr. 25 & 850.9  &  $-$14.6 & Du~Pont & WFCCD  & 3800 -- 9235   & 8   & 600   \\
\enddata
\tablenotetext{a}{Days since $B_{\rm max}$.} 
\tablenotetext{b}{The spectrum was obtained by the combination of two spectra from two different grisms, respectively covering the range between 3785 and 6129~\AA\ and the one between 5673 and 9969~\AA. The resolutions were different, being 2~\AA\ on the blue side and 4~\AA\ on the red one.}
\end{deluxetable}

\clearpage
\begin{deluxetable}{ccccccccc}
\tablewidth{0pt}
\tabletypesize{\scriptsize}
\tablecaption{Photometric observations of SN~2006V in the optical filters.\label{tabphot06V}}
\tablehead{
\colhead{JD~-~$2,453,000$} &
\colhead{Phase\tablenotemark{a}} &
\colhead{$u$ (mag)}&
\colhead{$g$ (mag)}&
\colhead{$r$ (mag)}&
\colhead{$i$ (mag)}&
\colhead{$B$ (mag)}&
\colhead{$V$ (mag)}&
\colhead{Instrument}}
\startdata
771.17  &   $-$52.53&    \ldots    &   \ldots        &  18.0\tablenotemark{b}        &    \ldots     &   \ldots      & \ldots &  LOT CCD\\  
773.15  &   $-$50.55&    \ldots    &   \ldots        &  18.2\tablenotemark{b}        &    \ldots     &   \ldots      & \ldots  &  LOT CCD \\  
773.71  &   $-$49.99&21.933(492)   & 18.724(012)     &  18.076(009)  &  18.021(011)  &  19.306(024)  & 18.355(012) 	&	 Site 3 		 \\
774.82  &   $-$48.88&21.454(231)   & 18.712(011)     &  18.042(009)  &  17.980(011)  &  19.209(020)  & 18.305(012) & Site 3 \\
775.67  &   $-$48.03&21.059(191)   & 18.684(010)     &  18.002(009)  &  17.949(011)  &  19.171(020)  & 18.267(010) & Site 3 \\
778.81  &   $-$44.89&21.082(321)   & 18.546(019)     &  17.890(016)  &  17.825(015)  &  19.034(039)  & 18.128(019) & Site 3 \\
784.90  &   $-$38.80&19.858(380)   &   \ldots        & \ldots      &  \ldots         &  18.912(050)  & 17.956(016) & Site 3 \\
786.88  &   $-$36.82&20.559(218)   & 18.302(011)     &  17.657(009)  &  17.610(010)  &  18.772(021)  & 17.890(011) & Site 3 \\
795.84  &   $-$27.86&\ldots        & 18.092(006)     &  17.455(007)  &  17.414(008)  &  18.613(013)  & 17.701(007) & Site 3 \\
799.75  &   $-$23.95&\ldots        & 17.999(006)     &  17.374(006)  &  17.344(007)  &  18.537(011)  & 17.616(006) & Site 3 \\
804.84  &   $-$18.86&\ldots        & 17.916(006)     &  17.291(005)  &  17.252(006)  &  18.435(009)  & 17.532(007) & Site 3 \\
805.79  &   $-$17.91&\ldots        & 17.901(007)     &  17.282(007)  &  17.239(008)  &  18.405(013)  & 17.531(008) & Site 3 \\
818.77  &    $-$4.93&\ldots        & 17.769(009)     &  17.135(008)  &  17.096(009)  &  18.287(013)  & 17.388(009) & Site 3 \\
824.73  &    $+$1.03&\ldots        & 17.749(006)     &  17.106(006)  &  17.059(007)  &  18.291(009)  & 17.351(006) & Site 3 \\
832.75  &    $+$9.05&\ldots        & 17.813(006)     &  17.130(007)  &  17.071(008)  &  18.344(010)  & 17.391(007) & Site 3 \\
838.75  &   $+$15.05&\ldots        & 17.942(016)     &  17.205(009)  &  17.144(010)  &  18.458(041)  & 17.499(014) & Site 3 \\
846.73  &   $+$23.03&\ldots        & 18.258(008)     &  17.430(006)  &  17.339(008)  &  18.888(014)  & 17.787(007) & Site 3 \\
853.64  &   $+$29.94&\ldots        & 18.752(009)     &  17.795(008)  &  17.664(009)  &  19.426(018)  & 18.219(009) & Site 3 \\
862.57  &   $+$38.87&\ldots        & 19.263(016)     &  18.186(010)  &  18.041(012)  &  19.950(030)  & 18.676(014) & Site 3 \\
867.57  &   $+$43.87&\ldots        & 19.337(029)     &  18.278(012)  &  18.143(017)  &  20.174(086)  & 18.757(026) & Site 3 \\
892.56  &   $+$68.86&\ldots        & 19.697(026)     &  18.518(018)  &  18.443(024)  &  \ldots       & 19.052(048) &Tek 5\\
898.56  &   $+$74.86&\ldots        & 19.759(020)     &  18.567(015)  &  18.500(021)  &  20.468(034)  & 19.213(018) &Tek 5\\      
\enddata 
\tablecomments{Values in parentheses are 1$\sigma$ measurement uncertainties in
  $10^{-3}$ of mag.}
\tablenotetext{a}{Days since $B_{\rm max}$, $JD=2453823.7$.}
\tablenotetext{b}{Unfiltered magnitudes from discovery and confirmation images, obtained with the Lulin One-meter Telescope (LOT) in Taiwan \citep{chen06}.}
\end{deluxetable}

\begin{deluxetable}{cccccc}
\tablewidth{0pt}
\tabletypesize{\scriptsize}
\tablecaption{Photometric observations of SN~2006V in the NIR filters.\label{tabphotIR06V}}
\tablehead{
\colhead{JD~-~$2,453,000$} &
\colhead{Phase\tablenotemark{a}} &
\colhead{$Y$ (mag)}&
\colhead{$J$ (mag)}&
\colhead{$H$ (mag)}&
\colhead{Instrument}}
\startdata
777.78 &  $-$45.92  & 17.306(029)  & 17.053(048)  & 16.919(063)& Retrocam \\
779.79 &  $-$43.91  & 17.283(038)  & 17.002(039)  & 16.819(062)& Retrocam \\
788.86 &  $-$34.84  & 17.059(012)  & 16.862(018)  & 16.713(027)& WIRC     \\
797.75 &  $-$25.95  & 16.994(035)  & 16.659(035)  & 16.502(056)& RetroCam \\
803.84 &  $-$19.86  & 16.767(027)  & \ldots       & \ldots     & RetroCam \\
810.79 &  $-$12.91  & 16.648(024)  & 16.450(030)  & 16.245(047)& RetroCam \\
817.71 &   $-$5.99  & 16.600(027)  & 16.361(026)  & 16.087(024)& RetroCam \\
826.68 &   $+$2.98  & 16.538(019)  & 16.275(020)  & 16.096(031)& RetroCam \\
834.74 &  $+$11.04  & 16.569(010)  & 16.343(016)  & 16.193(030)& WIRC     \\
844.65 &  $+$20.95  & 16.723(012)  & 16.514(039)  & 16.261(026)& WIRC     \\
852.64 &  $+$28.94  & 16.971(027)  & 16.692(035)  & 16.450(056)& RetroCam \\
863.59 &  $+$39.89  & 17.261(050)  & 17.148(063)  & \ldots     & RetroCam \\
869.64 &  $+$45.94  & 17.428(038)  & 17.209(048)  & 16.853(061)& RetroCam \\
\enddata 
\tablecomments{Values in parentheses are 1$\sigma$ measurement uncertainties in
  $10^{-3}$ of mag.}
\tablenotetext{a}{Days since $B_{\rm max}$, $JD=2453823.7$.}
\end{deluxetable}

\clearpage
\begin{deluxetable}{ccccccccc}
\tablewidth{0pt}
\tabletypesize{\scriptsize}
\tablecaption{Photometric observations of SN~2006au in the optical filters.\label{tabphot06au}}
\tablehead{
\colhead{JD~-~$2,453,000$} &
\colhead{Phase\tablenotemark{a}} &
\colhead{$u$ (mag)}&
\colhead{$g$ (mag)}&
\colhead{$r$ (mag)}&
\colhead{$i$ (mag)}&
\colhead{$B$ (mag)}&
\colhead{$V$ (mag)}&
\colhead{Instrument}}
\startdata
801.70 & $-$63.80  &    \ldots    &   \ldots     & 17.2\tablenotemark{b}&\ldots &   \ldots    & \ldots     & AP-7 CCD\\  
805.15 & $-$60.35  &    \ldots    &   \ldots     & 17.4\tablenotemark{b}&\ldots &   \ldots    & \ldots     & AP-7 CCD\\  
805.89 & $-$59.61  & 20.609(210)  & 18.428(015)  & 17.614(027)  & 17.106(093)  & 19.079(018)  & 17.991(013)&Site 3 \\
809.85 & $-$55.65  & \ldots       & 18.366(027)  & 17.589(028)  & 17.327(026)  & 19.308(065)  & 17.963(020)&Site 3 \\
815.86 & $-$49.64  & 20.815(198)  & 18.468(024)  & 17.696(051)  & 17.388(039)  & 19.206(027)  & 18.061(024)&Site 3 \\
818.90 & $-$46.60  & 20.446(230)  & 18.563(016)  & 17.637(023)  & 17.335(028)  & 19.219(023)  & 18.079(019)&Site 3\\
819.90 & $-$45.60  & 20.751(225)  & 18.578(022)  & 17.575(035)  & \ldots       & 19.244(020)  & 18.056(020)&Site 3\\
823.83 & $-$41.67  & 20.123(287)  & 18.549(016)  & 17.604(024)  & 17.371(026)  & 19.144(018)  & 18.061(018)&Site 3\\
824.87 & $-$40.63  & 20.448(226)  & 18.485(011)  & 17.557(016)  & 17.369(018)  & 19.149(019)  & 18.036(013)&Site 3\\
828.89 & $-$36.61  & 20.555(265)  & 18.321(036)  & \ldots       & \ldots       & 19.054(010)  & 17.953(018)&Tek 5\\
829.92 & $-$35.58  & 19.982(237)  & 18.542(038)  & \ldots       & \ldots       & 19.056(025)  & \ldots     &Tek 5\\     
830.89 & $-$34.61  & 20.482(257)  & \ldots       & 17.463(071)  & 17.402(081)  & 19.011(013)  & 18.045(039)&Tek 5\\
831.83 & $-$33.67  & 20.288(299)  & 18.304(018)  & 17.451(018)  & 17.362(025)  & \ldots       & 17.982(018)&Site 3\\
832.86 & $-$32.64  & 20.389(297)  & 18.408(009)  & \ldots       & \ldots       & 18.992(015)  & 17.957(010)&Site 3\\
835.87 & $-$29.63  & 20.285(282)  & 18.335(021)  & 17.501(037)  & 17.239(025)  & 18.923(016)  & 17.918(018)&Site 3\\
838.82 & $-$26.68  & 20.291(314)  & 18.175(021)  & 17.376(024)  & 17.083(033)  & 18.775(040)  & 17.775(018)&Site 3\\
840.83 & $-$24.67  & 20.185(224)  & 18.121(024)  & 17.376(025)  & 17.168(024)  & 18.796(029)  & 17.760(015)&Site 3\\
841.85 & $-$23.65  & 19.820(225)  & 18.133(021)  & 17.341(029)  & 17.112(023)  & 18.788(026)  & 17.749(019)&Site 3\\
845.82 & $-$19.68  & \ldots       & 18.096(012)  & 17.266(017)  & 17.043(019)  & 18.703(021)  & 17.666(012)&Site 3\\
850.81 & $-$14.69  & \ldots       & 18.087(011)  & 17.222(017)  & 16.991(020)  & 18.720(014)  & 17.601(010)&Site 3\\
853.85 & $-$11.65  & \ldots       & 18.056(016)  & 17.185(025)  & 16.974(023)  & 18.677(013)  & 17.598(016)&Site 3\\
858.79 & $-$6.71  & \ldots       & 18.016(007)  & 17.105(006)  & 16.929(015)  & 18.649(020)  & 17.506(015)&Site 3\\
861.82 & $-$3.68  & \ldots       & 18.006(016)  & 17.099(028)  & 16.867(023)  & 18.654(013)  & 17.535(016)&Site 3\\
862.80 & $-$2.70  & \ldots       & 18.018(009)  & 17.084(008)  & \ldots       & 18.653(014)  & 17.509(009)&Site 3\\
866.75 & $+$1.25  & \ldots       & 17.966(024)  & 16.996(031)  & 16.824(020)  & 18.628(021)  & 17.464(017)&Site 3\\
867.78 & $+$2.28  & \ldots       & 17.978(029)  & 16.981(035)  & 16.866(035)  & 18.678(028)  & 17.491(024)&Site 3\\
870.80 & $+$5.30  & \ldots       & 18.028(018)  & 17.013(018)  & 16.764(017)  & 18.675(034)  & 17.467(012)&Site 3\\
871.79 & $+$6.29  & \ldots       & 18.031(017)  & 17.022(016)  & 16.799(018)  & 18.768(029)  & 17.486(015)&Site 3\\
872.78 & $+$7.28  & \ldots       & 18.085(030)  & 16.994(043)  & 16.792(032)  & 18.776(022)  & 17.562(026)&Site 3\\
886.78 & $+$21.28  & \ldots       & \ldots       & \ldots       & 17.049(059)  & 19.360(020)  & 18.041(041)&Tek 5\\
890.72 & $+$25.22  & \ldots       & 18.832(026)  & 17.469(024)  & 17.258(025)  & 19.437(179)  & 18.127(063)&Site 3\\
891.74 & $+$26.24  & \ldots       & 18.883(017)  & 17.551(023)  & 17.258(025)  & 19.692(026)  & 18.211(016)&Site 3\\
892.76 & $+$27.26  & \ldots       & 18.987(064)  & 17.623(069)  & 17.353(054)  & 19.729(013)  & 18.212(024)&Tek 5\\
893.71 & $+$28.21  & \ldots       & 19.097(073)  & 17.479(042)  & 17.305(044)  & 19.839(024)  & 18.299(016)&Tek 5\\
894.75 & $+$29.25  & \ldots       & \ldots       & 17.669(017)  & 17.473(019)  & 19.854(035)  & 18.340(010)&Tek 5\\
897.66 & $+$32.16  & \ldots       & \ldots       & 18.015(150)  & \ldots       & 20.477(095)  & 18.830(097)&Tek 5\\
898.71 & $+$33.21  & \ldots       & \ldots       & 18.434(116)  & \ldots       & 21.185(231)  & 19.041(202)&Tek 5\\
\enddata 
\tablecomments{Values in parentheses are 1$\sigma$ measurement uncertainties in
  $10^{-3}$ of mag.}
\tablenotetext{a}{Days since $B_{\rm max}$, $JD=2453865.5$.}
\tablenotetext{b}{Unfiltered magnitudes from discovery and confirmation images, obtained with the 0.35~m Tenagra telescope in Oslo \citep{trondal06}.}
\end{deluxetable}

\begin{deluxetable}{cccccc}
\tablewidth{0pt}
\tabletypesize{\scriptsize}
\tablecaption{Photometric observations of SN~2006au in the NIR filters.\label{tabphotIR06au}}
\tablehead{
\colhead{JD~-~$2,453,000$} &
\colhead{Phase\tablenotemark{a}} &
\colhead{$Y$ (mag)}&
\colhead{$J$ (mag)}&
\colhead{$H$ (mag)}&
\colhead{Instrument}}
\startdata
807.87 &  $-$57.63 &  16.630(020)  & 16.398(025)  & 16.257(046)&RetroCam\\
808.85 &  $-$56.65 &  16.704(024)  & 16.375(027)  & 16.224(042)&RetroCam\\
810.87 &  $-$54.63 &  16.711(022)  & 16.377(020)  & \ldots     &RetroCam\\
813.87 &  $-$51.63 &  16.645(021)  & 16.380(019)  & \ldots     &RetroCam\\
820.87 &  $-$44.63 &  16.632(022)  & 16.371(018)  & 16.201(044)&RetroCam\\
826.83 &  $-$38.67 &  16.623(020)  & 16.366(022)  & 16.160(051)&RetroCam\\
834.87 &  $-$30.63 &  16.554(009)  & 16.248(009)  & 16.050(014)&WIRC\\
839.83 &  $-$25.67 &  16.456(019)  & 16.188(020)  & 16.050(040)&RetroCam\\
845.83 &  $-$19.67 &  16.353(008)  & 16.093(009)  & 15.838(013)&WIRC\\
848.83 &  $-$16.67 &  16.314(016)  & 16.052(016)  & 15.838(032)&RetroCam\\
868.80 &  $+$3.30 &  16.073(015)  & 15.794(011)  & 15.615(027)&RetroCam\\
\enddata 
\tablecomments{Values in parentheses are 1$\sigma$ measurement uncertainties in
  $10^{-3}$ of mag.}
\tablenotetext{a}{Days since $B_{\rm max}$, $JD=2453865.5$.}
\end{deluxetable}

\clearpage
\begin{deluxetable}{cccc|cccc}
\tablewidth{0pt}
\tabletypesize{\scriptsize}
\tablecaption{Absolute peak magnitudes of SN~2006V and SN~2006au in both optical and NIR filters.\label{absphotPEAK}}
\tablehead{
\colhead{Filter}&
\colhead{JD~-~$2,453,000$}&
\colhead{Phase\tablenotemark{a}}&
\colhead{Peak (mag)}&
\colhead{JD~-~$2,453,000$}&
\colhead{Phase\tablenotemark{a}}&
\colhead{Peak (mag)}}
\startdata
& & {\bf SN~2006V}& & &{\bf SN~2006au}& \\
$g$  & 824.5 & 76.5 & $-$16.67 & 866.9& 72.9& $-$16.55 \\
$r$  & 825.0 & 77.0 & $-$17.28 & 867.5& 73.5& $-$17.20 \\
$i$  & 825.3 & 77.3 & $-$17.31 & 868.5& 74.5& $-$17.15 \\
$B$  & 823.7 & 75.7 & $-$16.14 & 865.5& 71.5& $-$16.02 \\
$V$  & 824.8 & 76.8 & $-$17.05 & 867.0& 73.0& $-$16.87 \\
$Y$  & 825.5 & 77.5 & $-$17.81 & 868.5& 74.5& $-$17.63 \\
$J$  & 825.8 & 77.8 & $-$18.06 & 868.5& 74.5& $-$17.81 \\
$H$  & 823.7 & 75.7 & $-$18.26 & 868.5& 74.5& $-$17.90 \\
\enddata 
\tablecomments{The largest uncertainty affecting the absolute magnitudes is related to the distance estimate, whose error is about $7\%$.}
\tablenotetext{a}{Days since explosion, $JD=2453748$ for \V\ and $JD=2453794$ for \au.}
\end{deluxetable}

\begin{deluxetable}{ccccccc}
\tablewidth{0pt}
\tabletypesize{\scriptsize}
\tablecaption{Physical parameters for BSG supernovae. \label{param}}
\tablehead{
\colhead{Parameter} &
\colhead{\object{SN~2006V}} &
\colhead{\object{SN~2006au}}&
\colhead{\object{SN~1987A}}&
\colhead{\object{SN~1998A}}&
\colhead{\object{SN~2000cb}}&
\colhead{\object{SN~2009E}}}
\startdata
Explosion energy ($10^{51}$~erg)  &  2.4   & 3.2          & 1.1   & 5.6       & 4.4   &  0.6   \\
Ejecta mass ($\Msun$)             &  17.0  & 19.3         & 11.8  & 22.0      & 22.3  &  19.0  \\
Radius ($\Rsun$)                  &  75    & 90           & 33    & $\leq$~86 & 35    &  101   \\
$\Ni$ mass ($\Msun$)              &  0.127 & $\leq$~0.073 & 0.078 & 0.11      & 0.083 &  0.039 \\
\enddata 
\tablecomments{SN~1998A data are from \citet{pastorello05}, SN~2000cb from \citet{utrobin11} and SN~2009E from \citet{pastorello11}. The parameters
 for our objects and SN~1987A have been obtained from the semi-analytic model explained in the text.}
\end{deluxetable}

\clearpage

\clearpage

\clearpage

\clearpage

\clearpage

\clearpage

\clearpage

\clearpage

\clearpage

\clearpage

\clearpage

\clearpage

\clearpage

\clearpage

\clearpage

\end{document}